\documentclass[final,5p,times,twocolumn]{elsarticle}

\usepackage{graphicx}

\usepackage{amssymb}

\journal{ArXiv}

\begin{document}

\begin{frontmatter}



\title{A general tool for LTE thermochemistry for adiabatic nondiffusive fluid dynamics: applications to 2D planar discontinuity flows in SPH}


\author{G. Lanzafame}

\ead{glanzafame@oact.inaf.it}
\ead{tel. +39-0957332316; fax +39-095330592}

\address{INAF - Osservatorio Astrofisico di Catania, Via S. Sofia
              78 - 95123 Catania, Italy}

\begin{abstract}
Chemical reactions in fluid dynamics deeply modify the flow physical conditions through both the contribution of the energy of reactions and the variation of the mean molecular weight and of the ratio of specific heats. This occurs typically on time scales largely much smaller than diffusive time scales of the produced chemicals, especially for shock waves coming from explosive events. In this work we show how it is possible to include a standing alone algorithm, dealing with both molecular and nuclear thermochemistry in the computational nondiffusive adiabatic flow dynamics in local thermal equilibrium (LTE) in an explicit scheme of integration of fluid dynamics equations, free of the adopted computational framework. In this paper, working in the Free Lagrangian GASPHER framework, belonging to the smooth particle hydrodynamics methods (SPH), some comparisons are made for planar discontinuity flows among reactive to the respective unreactive flow models, assuming the same initial physical conditions and simple chemical composition. Results show the importance of the role not only of the thermochemical reaction energy, but also of the mean molecular weight and of the ratio of specific heats.
\end{abstract}


\end{frontmatter}

\section{Introduction}
\label{}

  Chemical reactions, essential in the transformation of matter, are fundamental in the flow thermodynamics because of the energy contribution to the energy balance, as well as due to the variation of the mean molecular weight and of the ratio of specific heats.

  Reactive flows are a more and more challenging theme for the wide fields of applications, because of their multidisciplinary content regarding real flows. Some textbooks have been written in the scientific literature \citep{c1,c2,c3,c4,c5} on the theoretical and experimental aspects, with a particular emphasis on the atmospheric gases \citep{c6} and interstellar plasmas \citep{c7}, as well as on a huge multitude of industrial and laboratory applications \citep{c8}. However, the numerical algorithm description on how to implement a standing alone chemistry tool, both molecular and nuclear, in a fluid dynamics numerical code is an effort not yet accomplished up to now.

  Therefore, the main aim of this paper is to give a detailed description on the standing alone algorithm we built-up, relative both to gas phase thermochemical and thermonuclear processes in those numerical codes in which a Predictor-Evaluator-Corrector (PEC), or a Predictor-Evaluator-Corrector-Evaluator (PECE), numerical integration technique is used. In particular, this effort is here accomplished in the Smoothed Particle Hydrodynamics (SPH) framework \citep{c9}, and particularly in its GASPHER version \citep{c10}.

  The reactive Euler system of equations, as well as its numerical conversion in the SPH Lagrangian framework refer to a nondiffusive, adiabatic LTE fluid dynamics at the moment. The algorithm we propose could be implemented in a wider scenario because those standing alone sections of the code, dealing with thermochemical reactions, strictly concern with the fluid dynamic code sections, only adding their time derivative contributions to those relative to the density and to the energy equations of fluid dynamics. Therefore, both the unreactive fluid dynamics spatial derivatives and the time integration algorithms stay unchanged.

  Nondiffusive LTE adiabatic reactive flows are meaningful in that reactive fluid dynamics, where diffusive time scales are normally much longer than reaction and dynamic time scales. Any chemical quenching of reactions, due to the flow cooling (either radiative or because of the adiabatic expansion), and the successive diffusive transport of reaction products deal with other themes of fluid dynamics, here not considered.

  To this purpose, in \S 2 of this paper we recall the system of equations to be solved. Although the new standing alone chemistry does not concern with a specific scheme of description of fluid dynamics, in the same section we describe how a PEC-PECE GASPHER technique works. In \S3 we formulate the rate of the local reactive mass density for gases both for chemical and for nuclear reactive flows and we explain how the algorithm is built-up and where it is placed within a PEC or a PECE integration code. In the same section we also show how the energy rate, produced by the thermochemical processes, is calculated. In \S4, we discuss some flow charts showing the algorithm structure to the chemistry contribution and how it is merged into the unreactive flow dynamics code. In \S5 we show results relative to 2D planar discontinuity flow structures and to their dynamic evolutions, considering either molecular or nuclear thermochemical reactions in the fluid. The comparison of each reactive discontinuity flow profile is discussed with respect to that relative to the simpler unreactive modelling, adopting the same initial conditions, limiting our results to a small number or reactions, because in so doing it is better possible the understanding of the role of physical parameters within the assumed State equation (EoS). Conclusions and a short discussion of results of simulations are reported in \S6, also considering the existence of other numerical reactive fluid dynamics codes, working with plasma nuclear thermochemistry flows.

\section{Adiabatic nonviscous reactive flows in LTE: how a general formulation is included into the GASPHER approach}
\label{}

\subsection{The equations of adiabatic nonviscous reactive flows in LTE}

  In the physically adiabatic, nondiffusive non viscous flows, the hyperbolic Euler system of equations in the Lagrangian scheme in LTE for reactive flows:
  
\begin{equation}
\frac{d \rho}{dt} + \rho \nabla \cdot \mathbf{v} = \sum_{k=1, r=1}^{K, R} \frac{d \rho_{k}}{dt} \Big|_{r} \hspace{.5 cm} \mbox{continuity equation}
\end{equation}

\begin{equation}
\frac{d \mathbf{v}}{dt} = - \frac{\nabla p}{\rho} + \mathbf{f} \hspace{2.2 cm} \mbox{momentum equation}
\end{equation}

\begin{eqnarray}
\frac{d}{dt} \left( \epsilon + \frac{1}{2} v^{2}\right) = - \frac{1}{\rho} \nabla \cdot \left( p \mathbf{v} \right) + \mathbf{f} \cdot \mathbf{v} + \sum_{k=1, r=1}^{K, R} \frac{d \epsilon_{k}}{dt} \Big|_{r} \nonumber
\end{eqnarray}

\begin{equation}
\hspace{4.3 cm} \mbox{energy equation}
\end{equation}

\begin{equation}
\frac{d \mathbf{r}}{dt} = \mathbf{v} \hspace{3.25 cm} \mbox{kinematic equation}
\end{equation}

  must be solved, together with the state equation (EoS) of the fluid
  
\begin{equation}
p = f(\gamma, \rho, \epsilon, \mathbf{r}, \mathbf{v}) \hspace{1.8 cm} \mbox{state equation.}
\end{equation}

$k = 1, ..., K$ is an index relative to the specific chemical species among the totality of $K$ chemicals, and $r = 1, ..., R$ is an index relative to the rth chemical reaction. $d/dt$ stands for the Lagrangian derivative, $\rho$ is the total gas mass density, $\epsilon$ is the thermal energy per unit mass, $p$ is the gas pressure, here generally expressed as a function $f(...)$ of local properties, $\mathbf{v}$ and $\mathbf{r}$ are the vectors velocity and position, $\mathbf{f}$ is the external force field per unit mass. The adiabatic index $\gamma = c_{p} c_{V}^{-1}$ is the ratio of specific heats.

\begin{equation}
\sum_{k=1, r=1}^{K, R} \frac{d \rho_{k}}{dt} \Big|_{r} = 0
\end{equation}

for chemical molecular reactions, while

\begin{equation}
\sum_{k=1, r=1}^{K, R} \frac{d \rho_{k}}{dt} \Big|_{r} \neq 0
\end{equation}

for chemical nuclear reactions.

For nuclear reactions, the mass to the binding energy conversion (and vice versa) needs to be taken into account.

  In LTE conditions, chemical molecular reactions contribute only with the algebraic summation term to the right side of the energy equation, while the continuity equation could be involved for nuclear reactions. For each chemical species $k$, the continuity equation is:

\begin{equation}
\frac{d \rho_{k}}{dt} + \rho_{k} \nabla \cdot \mathbf{v} = \sum_{r=1}^{R} \frac{d \rho_{k}}{dt} \Big|_{r},
\end{equation}

being $d \rho_{k}/dt|_{r}$ the explicit algebraic contribution due to the $r$th reaction to the $k$th chemical component. Moreover, $\sum_{k=1}^{K} \rho_{k} = \rho$. In so doing, the summation over $k$ of eq. (8) gives exactly eq. (1). If $\varepsilon_{k}$ is the energy contribution per unit mass corresponding to the $k$th chemical species (e.g. its formation enthalpy $\Delta H^{0}_{f, k}$ or its rest mass-energy $q_{k}$), for the rth reaction

\begin{equation}
\frac{d \epsilon_{k}}{dt} \Big|_{r} = - \frac{\varepsilon_{k}}{\rho} \frac{d \rho_{k}}{dt} \Big|_{r}.
\end{equation}

Notice that $\epsilon_{k}$ and $\varepsilon_{k}$ do not conceptually equal with each other, the first being a computed energy per unit mass, while the second being an assigned energy per unit mass for the same $k$th chemical species.

  It is important to note that the temporal variations due only to the reaction rates can indifferently be either temporal Lagrangian derivatives $d/dt$ or temporal Eulerian derivatives $\partial/\partial t$, being any chemical variation strictly local. Henceforth, we will indistinctly adopt the $\dot{A}$ formalism to identify the reaction rate of the chemical species $A$, instead of $\partial A/\partial t$ or of $dA/dt$. This implies that any standing alone numerical algorithm, written for the calculation of the flow chemical composition could directly be implemented in any numerical scheme. In that regard, we adopt a finite volumes Lagrangian scheme of moving smooth domains, based on the well known SPH framework \citep{c9}, built-up on the basis of the mathematical error function as a profile of the spatial distribution of the smooth physical properties \citep{c10}.

\subsection{The GASPHER interpolation Kernel in the SPH framework}

  The SPH method is a nonlinear Free Lagrangian scheme \citep{c11} depicting the fluid into interacting and interpolating domains called "particles" \citep{c12,c13}, moving according to pressure and body forces. The method makes use of a Kernel $W$ useful to smoothing interpolate a physical quantity $Q(\mathbf{r})$ related to a gas particle at position $\mathbf{r}$ according to:

\begin{equation}
\overline{Q}(\mathbf{r}) = \int_{D} Q(\mathbf{r}') W(\mathbf{r}, \mathbf{r}', h) d \mathbf{r}'.
\end{equation}

$W(\mathbf{r}, \mathbf{r}', h)$, the interpolation Kernel, is a continuous function - or two or more connecting continuously differentiable functions at their connecting point - defined within a spatial window, whose spatial resolution length limit for $h \rightarrow 0$ is the Dirac delta distribution function. All physical quantities are described as extensive properties, smoothly distributed in space, and computed by interpolation at $\mathbf{r}$. Therefore, in the SPH formalism:

\begin{equation}
\overline{Q}(\mathbf{r}_{i}) = \overline{Q}_{i} = \sum_{j=1}^{N} \frac{Q_{j}}{n_{j}} W(\mathbf{r}_{i}, \mathbf{r}_{j}, h) = \sum_{j=1}^{N} \frac{Q_{j}}{n_{j}} W_{ij}
\end{equation}

at the position $\mathbf{r}_{i}$ of the $i$th particle. The sum is extended to all neighbour particles included within the domain $D$, $n_{j} = \rho_{j} m_{j}^{-1}$ is the number density relative to the jth particle. $W(\mathbf{r}_{i}, \mathbf{r}_{j}, h)$ is the adopted interpolation Kernel, whose value is determined by the relative distance between particles $i$ and $j$. $\int W(\mathbf{r}_{i}, \mathbf{r}_{j}, h) d^{3} \mathbf{r}' = 1$, that is: $\sum_{j} W(\mathbf{r}_{i}, \mathbf{r}_{j}, h) n_{j}^{-1} = 1$.

  In the SPH mathematical equations there are two principles embedded. Each SPH particle is an extended domain, spherical or lenticular. Within that domain, any physical quantity $Q$ has a density profile $Q W(\mathbf{r}_{i}, \mathbf{r}_{j}, h) \equiv Q W(|\mathbf{r}_{i} - \mathbf{r}_{j}|, h) = Q W(|\mathbf{r}_{ij}|,h)$. Besides, the fluid quantity $\overline{Q}$ at the position of each SPH particle could be interpreted by filtering the particle data for $Q(\mathbf{r})$ with a single windowing function whose width is $h$. In so doing, fluid physical properties are considered isotropically smoothed all around each particle along a length scale $h$. Therefore, according to these two concepts, the SPH value $\overline{Q}$ of the physical quantity $Q$ is both the overlapping of extended profiles of all particles and the overlapping of the closest smooth density profiles of $\overline{Q}$. This means that the compactness of the Kernel shape gives the principal contribution to the interpolation summation to each particle by itself and by its closest neighbours. In both approaches the mass of particles is globally conserved in so far as it is the real physical mass.

  In the SPH formalism, equations (1-3) take the form:
  
\begin{eqnarray}
\frac{d \rho_{i}}{dt} & = & - \sum_{j=1}^{N} \frac{m_{j}}{\rho_{j}} \mathbf{v}_{ij} \cdot \nabla_{i} W_{ij} + \sum_{k=1, r=1}^{K, R} \frac{d \rho}{dt} \Big|_{k, r} \\
\frac{d \mathbf{v}_{i}}{dt} & = & - \sum_{j=1}^{N} m_{j} 
\left( \frac{p_{i}^{\ast}}{\rho_{i}^{2}} + \frac{p_{j}^{\ast}}{\rho_{j}^{2}} \right) \nabla_{i} W_{ij} + \mathbf{f}_{i} \\
\frac{d}{dt} E_{i} & = & - \sum_{j=1}^{N} m_{j} \left( \frac{p_{i}^{\ast} \mathbf{v}_{i}}{\rho_{i}^{2}} + \frac{p_{j}^{\ast} \mathbf{v}_{j}}{\rho_{j}^{2}}\right) \cdot \nabla_{i} W_{ij} + \nonumber \\ & & \sum_{k=1, r=1}^{K, R} \frac{d \epsilon}{dt} \Big|_{k, r}
\end{eqnarray}

where $\mathbf{v}_{ij} = \mathbf{v}_{i} - \mathbf{v}_{j}$, $m_{j}$ is the mass of $j$th particle, $p_{i}^{\ast} = p_{i} +$ {\it artificial pressure term} and $E_{i} = \epsilon_{i} + v_{i}^{2}/2$.

  Notice that the alternative SPH conversion of the fluid component of the continuity equation (eq. 1) as:

\begin{equation}
\rho_{i} = \sum_{j=1}^{N} m_{j} W_{ij}
\end{equation}

which identifies the natural space interpolation of particle densities, according to equations (8-9), does not directly apply if, as for nuclear reactive flows, further terms could be included in the continuity equation as a consequence of temporal local variations of the fluid chemical composition.

  A thermal conductivity numerical term $q_{ij} \propto h_{ij} c_{sij} \rho_{ij}^{-1} (\epsilon_{i} - \epsilon_{j}) (r_{i} - r_{j})^{-1}$ is usually added to the pressure terms within the parenthesis in the first summation in the energy equation (eq. 14), useful to smooth out spurious discontinuities in the numerical solutions \citep{c9}. Any subscript $ij$ still means the average value calculated between the two $i$ - $j$ particles.

  Several Kernel formulations exist in the literature (e.g. \citep{c14,c15,c16,c17}). However, it is a rather common weak side of most of SPH Kernels the problem of particle pairing instability whenever the mutual distance between two flow particles $r_{ij} \ll h$ \citep{c10}.

  In GASPHER modelling \citep{c10}, a radial Gaussian-derived Kernel, related to the well-known "Error Function" is considered:

\begin{equation}
\left\{ \begin{array}{llll}
W_{ErF,ij} = \frac{2}{\pi^{1/2} h} e^{- r_{ij}^{2}/h^{2}} \nonumber \hspace {1.45cm} \mbox{in 1D,} \\
W_{ErF,ij} = \frac{1}{\pi^{3/2} h r_{ij}} e^{- r_{ij}^{2}/h^{2}} \hspace {1.22cm} \mbox{in 2D,} \\
W_{ErF,ij} = \frac{1}{2 \pi^{3/2} h r_{ij}^{2}} e^{- r_{ij}^{2}/h^{2}}. \nonumber \hspace {1cm} \mbox{in 3D,} \\
\end{array} \right.
\end{equation}

  with a constant smoothing length $h$ equivalent to its half width at half maximum. In these Kernel formulations, it is important to note that, if $d$ is the number of dimensions, the term $h r^{d-1}_{ij}$ substitutes the original $h^{d}$ strictly used for Gaussian formaulations, and that the same $h r^{d-1}_{ij} = h^{d} (r_{ij} h^{-1})^{d-1}$. This means that not only $r^{d-1}_{ij}$ is used instead of $h^{d-1}$, but also that the entire $(r_{ij} h^{-1})^{d-1}$ term is necessary since the integration itself introduces the same multiplier terms $2 \pi$ or $4 \pi$ in the continuum limit. In doing so, $W_{ErF,ij} d\mathbf{r}_{ij}$, as well as $W_{ErF,ij} n^{-1}_{j}$ in the continuum limit, are invariant on the number of dimensions. The interpolation radial extension is theoretically unlimited although the smoothing resolution length $h$ is constant. This means that the problem of a small number of particle neighbours does not exist.

  The origin of this Kernel function relies in the well known "Error Function":

\begin{equation}
ErF(x) = \frac{2}{\sqrt{\pi}} \int_{0}^{x} e^{- t^{2}} dt,
\end{equation}

whose "Complementary Error Function" is:

\begin{equation}
ErFC(x) = 1 - ErF(x) = \frac{2}{\sqrt{\pi}} \int_{x}^{\infty} e^{- t^{2}} dt.
\end{equation}

  For $x = 0$,

\begin{equation}
ErFC(0) = 1 - ErF(0) = \frac{2}{\sqrt{\pi}} \int_{0}^{\infty} e^{- t^{2}} dt = 1.
\end{equation}

  For $x = 0$, $ErFC(0)$ equals the zero order Gaussian integral:

\begin{equation}
I_{0} = \int_{0}^{\infty} e^{- \xi t^{2}} dt = \frac{1}{2} \frac{\sqrt{\pi}}{\xi}.
\end{equation}

  In performing 3D integral,
  
\begin{eqnarray}
\int W_{ErF,ij} d^{3} r_{ij} & = & 4 \pi \int_{0}^{\infty} W_{ErF,ij} r_{ij}^{2} dr_{ij} \nonumber \\ & = & 4 \pi \int_{0}^{\infty} \frac{1}{2 \pi^{3/2} h r_{ij}^{2}} e^{- r_{ij}^{2}/h^{2}} r_{ij}^{2} dr_{ij} \nonumber \\ & = & 4 \pi \int_{0}^{\infty} \frac{1}{2 \pi^{3/2} h} e^{- r_{ij}^{2}/h^{2}} dr_{ij} \nonumber \\ & = & \frac{2}{\sqrt{\pi}} \int_{0}^{\infty} e^{- q_{ij}^{2}} dq_{ij} = 1,
\end{eqnarray}

where, $q_{ij} = |\mathbf{r}_{ij}|/h$.
  
  It is worth noting that the integral of the Gauss function: $\int W_{G,ij} d^{3} r_{ij} = 1$. This according to the well known properties of Gaussian integrals: $I_{n} = \int_{0}^{\infty} t^{n} e^{- \xi t^{2}} dt$. In particular $I_{2} = \int_{0}^{\infty} t^{2} e^{- t^{2}} dt = \pi^{1/2}/4$ could be interesting for a comparison with the 3D formulation for the interpolation integral.

  Two things are relevant:

\begin{itemize}

\item according to eqs. (10, 11) $\int_{D} W(\mathbf{r}, \mathbf{r}', h) d \mathbf{r}' = 1$, that means that $\sum_{j=1}^{N} n^{-1}_{j} W_{ij} = 1$, a condition required for SPH-like Kernels, that is mathematically fulfilled by using eq. (16). It is worth noting that, in 2D and in 3D, the function to be integrated it is not only the Kernel, but also the contribution coming from $d \mathbf{r}'$ (pertaining to $n^{-1}_{j}$ in the summation - eqs. 10, 11), involving $2 \pi r$ and $4 \pi r^{2}$ terms, respectively. Besides, any $- \nabla W_{ij} \rightarrow \infty$ as $r_{ij} \rightarrow 0$ so that pressure forces physically go to infinity whenever the mutual separation of two SPH particles goes to zero. Any situation of pressure forces numerically too high could occur only in extreme accretion conditions in 2D and in 3D, where the particle merging option could be a practical solution to any numerical shortcoming.

\item Even though the particle resolution length $h$ is constant, a number of assigned companions can be reached, thanks to the fact that the Kernel extension is theoretically unlimited, a particular that is largely used in numerous 1D SPH exponential Kernels \citep{c14, c15, c16, c17}. Using these form of Kernels, all companions within a distance of $2 h$ are considered. However if their number is smaller than the assigned minimum number of companions within $12 - 15$ in 2D ($40 - 50$ in 3D), the counting of further farther neighbours continues up to the wished number of companions to get a satisfactory accuracy of interpolations. In so doing, any discontinuity at the outer boundary (if truncating) is non-existing. Further contributions, coming from other farther particles, are almost negligible and smaller than the conversion errors coming from the numerical summations instead of strictly mathematical integrals \citep{c18,c19}, a particular also affecting any traditional SPH interpolation technique, working with finite extension Kernels, but suffering of particle pairing instabilities \citep{c10}.

\end{itemize}

  Comparisons of analytical solutions with numerical GASPHER results are in very good agreement and shown in \citep{c10}, so that another comparison here is useless. Besides, the GASPHER Kernel integration also solves problems related to free edge boundaries. As for the computational cpu time, there is not conceptually any disadvantage in such a transformation, being the number of particle neighbours assigned (e.g. $30$ or $50$) for each particle, by the introduction of a boundaries counter/limiter, since the number of particle neighbours rules the computational cpu time.

  Hydrodynamics in the nonlinear Free Lagrangian SPH approach is currently performed in predictor-corrector explicit schemes, starting from some initial values at time $t = 0$. In the "Leapfrog" scheme \citep{c20,c21,c22}, the equations for space and velocity advancement can be written as:

\begin{eqnarray}
r_{l}^{n+1} & = & r_{l}^{n} + v_{l}^{n+1/2} \Delta t \\
v_{l}^{n+1/2} & = & v_{l}^{n-1/2} + a_{l}^{n} \Delta t
\end{eqnarray}

that can be manipulated into a form which writes particle velocity at integer steps as

\begin{eqnarray}
r_{l}^{n+1} & = & r_{l}^{n} + v_{l}^{n} \Delta t + \frac{1}{2} a_{l}^{n} \Delta t^{2} \\
v_{l}^{n+1} & = & v_{l}^{n} + \frac{1}{2} (a_{l}^{n} + a_{l}^{n+1}) \Delta t.
\end{eqnarray}

  In the second expression, since particle acceleration $a$ depends on $v$, it is required an implicit integration for the second equation. In the case of a "Leapfrog" scheme, an "Evaluator" phase in the computational scheme needs to be interposed between the two integration procedures, where time derivatives of the various physical quantities are computed. For this reason, this scheme is a so called PEC method, where a Predictor-Evaluator-Corrector procedure is followed by the updating of all integrated values.

  Iterative Runge-Kutta methods are also used, both explicit as well as implicit, like the simple backward Euler, or the Crank-Nicholson methods \citep{c20,c21,c22}. Despite being more general, implicit schemes are more complicated and dependent on the specific problem. Those Runge-Kutta methods that are diagonally implicit, show a strong stability allowing a significant increase in the time step limit, with respect to explicit methods of the same order. Here, we do not consider any implicit technique, where often some Jacobian matrices need to be inverted \citep{c20,c21,c22}.

  The explicit multistep Adams-Bashforth-Moulton \citep{c23,c24} PECE explicit integration scheme can also be adopted in SPH, where either the Adams-Bashforth

\begin{equation}
\left\{ \begin{array}{llll}
A_{l}^{n+1} = A_{l}^{n} + \Delta t \frac{\partial A_{l}^{n}}{\partial t} \\
A_{l}^{n+2} = A_{l}^{n+1} + \frac{\Delta t}{2} \Bigg( 3 \frac{\partial A_{l}^{n+1}}{\partial t} - \frac{\partial A_{l}^{n}}{\partial t} \Bigg) \\
A_{l}^{n+3} = A_{l}^{n+2} + \frac{\Delta t}{12} \Bigg( 23 \frac{\partial A_{l}^{n+2}}{\partial t} - 16 \frac{\partial A_{l}^{n+1}}{\partial t} + 5 \frac{\partial A_{l}^{n}}{\partial t} \Bigg) \\
A_{l}^{n+4} = A_{l}^{n+3} + \frac{\Delta t}{24} \Bigg( 55 \frac{\partial A_{l}^{n+3}}{\partial t} - 59 \frac{\partial A_{l}^{n+2}}{\partial t} + \\ 37 \frac{\partial A_{l}^{n+1}}{\partial t} - 9 \frac{\partial A_{l}^{n}}{\partial t} \Bigg) \\
A_{l}^{n+5} = A_{l}^{n+4} + \frac{\Delta t}{720} \Bigg( 1901 \frac{\partial A_{l}^{n+4}}{\partial t} - 2774 \frac{\partial A_{l}^{n+3}}{\partial t} + \\ 2616 \frac{\partial A_{l}^{n+2}}{\partial t} - 1274 \frac{\partial A_{l}^{n+1}}{\partial t} + 251 \frac{\partial A_{l}^{n}}{\partial t} \Bigg),
\end{array} \right.
\end{equation}

or the Adams-Moulton

\begin{equation}
\left\{ \begin{array}{llll}
A_{l}^{n} = A_{l}^{n-1} + \Delta t \frac{\partial A_{l}^{n}}{\partial t} \\
A_{l}^{n+1} = A_{l}^{n} + \frac{\Delta t}{2} \Bigg( \frac{\partial A_{l}^{n+1}}{\partial t} + \frac{\partial A_{l}^{n}}{\partial t} \Bigg) \\
A_{l}^{n+2} = A_{l}^{n+1} + \frac{\Delta t}{12} \Bigg( 5 \frac{\partial A_{l}^{n+2}}{\partial t} + 8 \frac{\partial A_{l}^{n+1}}{\partial t} - \frac{\partial A_{l}^{n}}{\partial t} \Bigg) \\
A_{l}^{n+3} = A_{l}^{n+2} + \frac{\Delta t}{24} \Bigg( 9 \frac{\partial A_{l}^{n+3}}{\partial t} + 19 \frac{\partial A_{l}^{n+2}}{\partial t} - \\ 5 \frac{\partial A_{l}^{n+1}}{\partial t} + \frac{\partial A_{l}^{n}}{\partial t} \Bigg) \\
A_{l}^{n+4} = A_{l}^{n+3} + \frac{\Delta t}{720} \Bigg( 251 \frac{\partial A_{l}^{n+4}}{\partial t} + \\ 646 \frac{\partial A_{l}^{n+3}}{\partial t} - 264 \frac{\partial A_{l}^{n+2}}{\partial t} + 106 \frac{\partial A_{l}^{n+1}}{\partial t} - 19 \frac{\partial A_{l}^{n}}{\partial t} \Bigg).
\end{array} \right.
\end{equation}

are used. In applying either the Adams-Bashforth or the Adams-Moulton techniques, up to the wished precision, previous derivatives for the same flow elements need to be conserved. Besides, a further evaluator procedure is considered at the end of the predictor-corrector integration scheme, giving a PECE technique.

\subsection{The gas chemical composition in GASPHER-SPH particle hydrodynamics}

  From the initial particle setting, each flow element (here every $i$th SPH particle) contains a gas mixture in which the relative abundance of the $k$th chemical species is

\begin{equation}
X_{i, k} = n_{i, k} n^{-1}_{i},
\end{equation}

where $n_{i, k}$ is the number density of the $k$th chemical species within particle $i$, and

\begin{equation}
{n_{i}} = \sum_{k=1}^{K} n_{i, k},
\end{equation}

giving

\begin{equation}
\sum_{k=1}^{K} X_{i, k} = 1.
\end{equation}

\begin{figure}
\resizebox{\hsize}{10 cm}{\includegraphics[clip=true]{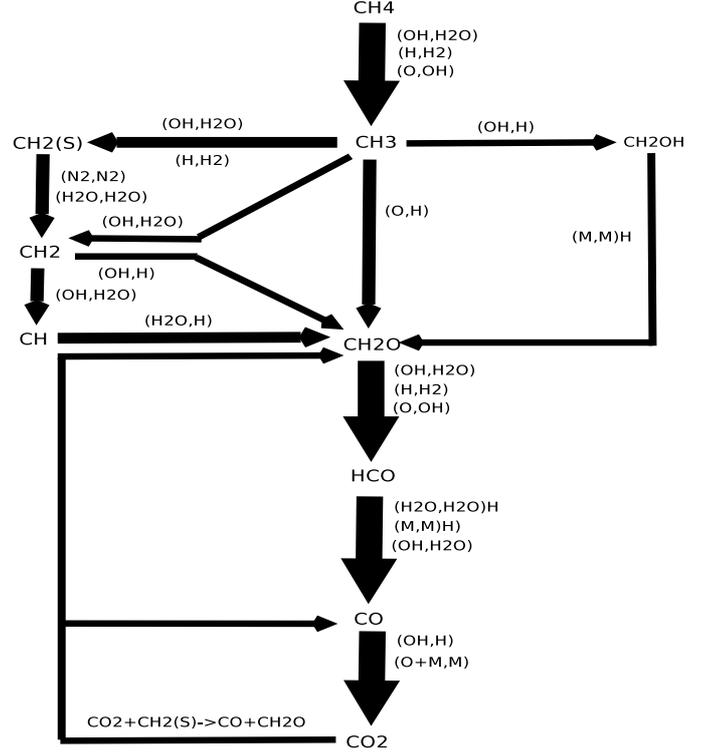}}
\caption{Scheme of the chain of methan combustion main reactions $CH_{4} + 2O_{2} \rightarrow CO_{2} + 2H_{2}O$, for $p = 1$ atm and $T = 2200$ K.}
\end{figure}

Being $\rho_{i} = m_{i} n_{i}$ and introducing the molecular weight $\mu_{k}$ of the $k$th species, since the mass density for the $k$th species for the $i$th particle

\begin{equation}
\rho_{i, k} = \mu_{k} N_{A}^{-1} n_{i, k} = \mu_{k} N_{A}^{-1} X_{i, k} n_{i},
\end{equation}

\begin{equation}
\rho_{i} = N_{A}^{-1} n_{i} \sum_{k=1}^{K} \mu_{k} X_{i, k} = \overline{\mu_{i}} N_{A}^{-1} n_{i},
\end{equation}

where

\begin{equation}
\overline{\mu_{i}} = \sum_{k=1}^{K} \mu_{k} X_{i, k}
\end{equation}

is the mean molecular weight and $N_{A}$ is the Avogadro's number. This means that in so doing the mass of a SPH particle should be strictly related to the real gas mass, instead of being an arbitrary parameter, useful only for particle interpolation as it is often currently made in nonreactive fluid dynamics.

\section{The rate of local mass density variation for reactions in the gas phase}
\label{}

\subsection{Molecular collisional rate for gas phase reactions}

  In a chemical reaction, collisions of some reactants give some reaction products, according to the general expression:

\begin{equation}
\sum_{k=1}^{K_{a}} s_{kr} A_{k} \rightarrow \sum_{k=1}^{K_{b}} s_{kr} B_{k},
\end{equation}

being $s_{kr}$ the stoichiometric coefficients for the $r$th reaction. $K_{a}$ and $K_{b}$ are the number of reactants and of reaction products, respectively. Whenever a reaction equilibrium occurs, the $\rightarrow$ in the expression (34) is replaced either by $=$ or by $\leftrightarrows$.

  Almost all reactions regarding free reactants involve binary collisions, even for reactions whose stoichiometric coefficients are greater than $1$. As an example, the well known methane combustion reaction:

\begin{equation}
CH_{4} + 2O_{2} \rightarrow CO_{2} + 2H_{2}O
\end{equation}

in reality involves a chain of reactions, according to the chain of main reactions at $T = 2200$ K shown in Fig. 1, where, for simplicity of representation the arbitrary chemical reaction $A + B \rightarrow C + D$ is represented as $A(B,C+D)$. Three colliding reactants are sometimes also considered, as shown in Fig. 1. However, in this case, the collision of the first two reactants leads to the formation of a transient reactant, whose lifetime is longer than the mean collision time of the third reactant \citep{c1}.

  The speed of a chemical reaction is proportional to the molar concentrations of the two reactants $[A] \propto n_{A}$ and $[B] \propto n_{B}$ as (here for the chemical species $A$):

\begin{equation}
\dot{[A]} = \dot{A}|_{r} = \frac{\partial [A]}{\partial t}\Big|_{r} = \frac{d [A]}{d t}\Big|_{r} = - C_{reac, AB}(T) [A]^{m}[B]^{n},
\end{equation}

being $n_{A}$ and $n_{B}$ the usual number densities in cm$^{-3}$ s$^{-1}$ and $C_{reac, AB} \propto \sigma v$ is the reaction rate coefficient, proportional to the reaction cross section $\sigma$ and to the molecular velocity $v$. Exponents $m$ and $n$, bear no relation to the stochiometric coefficients. They can be only determined experimentally and define the order of the chemical reactions $p = m + n + ...$. In most of the chemical combustion reactions, exponents $m$, $n = 1$. The reaction speed coefficient $C_{reac, AB}(T)$ is temperature dependent and should be experimentally determined. However, it is widely accepted that order $1$ reactions deal with molecular dissociation, nuclear decay or nuclear fission:

\begin{equation}
\left\{ \begin{array}{llll}
(AB) \rightarrow A + B \\
(ABC) \rightarrow (AB) + C \\
(ABC) \rightarrow A + B + C \\
etc.,
\end{array} \right.
\end{equation}

while $2$ and $3$ order chemical reactions deal with bimolecular collisions and "trimolecular'' collisions, respectively. Radiative molecular association (molecular recombination), as well as collision induced dissociations, where:

\begin{equation}
A + B \rightarrow (AB) + h \nu
\end{equation}

and

\begin{equation}
M + (AB) \rightarrow M + A + B,
\end{equation}

respectively, also belong to the group of order 2 collisional reaction processes.

  In a real gas, assuming a Maxwellian velocity distribution of all molecules, the collisional frequency between reactants $A$ and $B$ per unit volume is

\begin{equation}
\nu_{c_{AB}} = 2^{1/2} [A] [B] \sigma_{AB} \left( \overline{v}_{A}^{2} + \overline{v}_{B}^{2} \right)^{1/2},
\end{equation}

where $\sigma_{AB}$ is the collisional cross section and $\overline{v}_{A}$ and $\overline{v}_{B}$ are the mean velocity of specified molecules. This expression, as a function of temperature $T$, becomes:

\begin{equation}
\nu_{c_{AB}} = 2^{1/2} [A] [B] \sigma_{AB} \left( \frac{8 K_{B} T}{\pi m} \right)^{1/2},
\end{equation}

where $m = m_{A} m_{B} (m_{A} + m_{B})^{-1}$ is the reduced mass relative to the two reactants and $K_{B}$ is the Boltzmann constant. As a consequence, the reaction rate for reactant $A$ in $cm^{-3} s^{-1}$ units can be expressed as:

\begin{equation}
\dot{[A]} \propto - \nu_{c} P_{AB} N_{A}^{-1},
\end{equation}

being $P_{AB} < 1$ (sometimes $P_{AB} \ll 1$) the so called "steric factor" taking into account the collisional geometry between reactants $A$ and $B$. To consider those collisions effective for giving the reaction, we need to multiply expression (40) by the the energetic factor $e^{- E_{a}/RT}$, giving the probability that the collision gives the wished reaction. $R$ is the gas constant and $E_{a}$ is the activation energy, expressing the percentage of collisions whose energy is greater than the threshold energy needed for the reaction occurrence. Hence,

\begin{equation}
\dot{[A]} = - P_{AB} N_{A}^{-1} \sigma_{AB} \left( \frac{8 K_{B} T}{\pi m} \right)^{1/2} e^{- E_{a}/RT} [A] [B].
\end{equation}

  This result defines the reaction rate coefficient:

\begin{equation}
C_{reac, AB}(T) = P_{AB} N_{A}^{-1} \sigma_{AB} \left( \frac{8 K_{B} T}{\pi m} \right)^{1/2} e^{- E_{a}/RT},
\end{equation}

based on the collisional theory alone. Formulation (43) works whenever $A \neq B$. So that, a more general expression is:

\begin{equation}
\dot{[A]} = - P_{AB} N_{A}^{-1} \frac{\sigma_{AB}}{1 + \delta_{k_{A} k_{B}}} \left( \frac{8 K_{B} T}{\pi m} \right)^{1/2} e^{- E_{a}/RT} [A] [B],
\end{equation}

where $\delta_{k_{A} k_{B}}$ is the Kroneker delta.

  In the following, subscripts $A$, $B$, $AB$ for order 2 reactions, as well as subscript $(AB)$ for order 1 dissociation or decay of the chemical species before its breakup, are equivalent to the subscripts $k_{A}$, $k_{B}$, $k_{AB}$ and $k_{(AB)}$ counterparts, respectively, relative to the label (numerical index) within the nested do loops in the code.

  By adopting the numerical densities $n_{A}$ and $n_{B}$, instead of the molar densities $[A]$ and $[B]$, the analytical counterparts of eqs. (43 and 44) are respectively:

\begin{equation}
\dot{n_{A}} = - P_{AB} \frac{\sigma_{AB}}{1 + \delta_{k_{A} k_{B}}} \left( \frac{8 K_{B} T}{\pi m} \right)^{1/2} e^{- E_{a}/RT} n_{A} n_{B}
\end{equation}

and

\begin{equation}
C_{reac, AB}(T) = P_{AB}  \sigma_{AB} \left( \frac{8 K_{B} T}{\pi m} \right)^{1/2} e^{- E_{a}/RT}.
\end{equation}

  Unfortunately, the theory of collisions alone does not allow to find either the activation energy or the steric factor or both. Therefore, for the gas phase chemical reactions, it is considered the well known empirically modified Arrhenius formula:

\begin{equation}
C_{reac, AB}(T) = \alpha T^{\beta} e^{- \theta/T},
\end{equation}

that is valid within a certain temperature window, where $\alpha$, $\beta$ and $\theta$ are three empirical parameters.

\subsection{Molecular Dissociation}

  The empirically modified Arrhenius fitting formula (eq. 48) for the rate coefficients is also adopted for order 1 chemical reactions regarding the molecular dissociation processes. This is possible because the fitting parameter $\theta$ in eq. (48) equals the activation temperature $E_{a, k_{(AB)}} K_{B}^{-1}$ (or $E_{a, k_{(AB)}} R^{-1}$. Instead, the $\alpha T^{\beta}$ term in eq. (48) corresponds to the temperature dependent molecular dissociation coefficient $D_{k_{(AB)}}$ in s$^{-1}$, that is the constant of molecular dissociation, whenever $\beta = 0$. Consequently, for such order 1 reactions:

\begin{equation}
D_{k_{(AB)}} = \alpha T^{\beta},
\end{equation}

and

\begin{equation}
C_{reac, AB}(T) = C_{reac,(AB)}(T) = D_{k_{(AB)}} e^{- E_{a, k_{(AB)}}/K_{B} T},
\end{equation}

so that, the molecular dissociation rates, as well as the corresponding energy variations, are largely reduced for those reactants $(AB)$ having high values of $E_{a, k_{(AB)}}$ with respect to $K_{B} T$.

\subsection{Radiative and dielectronic recombination coefficients}

  Electron temperature $T_{e}$ for free electrons in the gas phase in LTE equals that relative to other barionic components since gas kinematic timescales are not shorter than atomic and molecular transition timescales in which a statistical equilibrium among the various ion populations is arranged. In the recombination reaction

\begin{equation}
A^{i+1} + e^{-} \rightarrow A^{i} + h \nu,
\end{equation}

if $n_{e}$ is the electronic number density, the recombination coefficient $\alpha_{rec} (A^{i+1})$ of an ion $A^{i}$ gives the reaction rate $\alpha_{rec} n^{A^{i+1}} n_{e}$, where $n^{A^{i+1}}$ is the number density of the reactant ion $A^{i+1}$ in the $i$th $+ 1$ initial ionization state. Such a recombination coefficient of $A^{i+1}$ is composed of two contributions:

\begin{equation}
\alpha_{rec} (A^{i+1}) = \alpha_{rad} (A^{i+1}) + \alpha_{di} (A^{i+1}),
\end{equation}

$\alpha_{rad} (A^{i+1})$ and $\alpha_{di} (A^{i+1})$ being the radiative and the dielectronic recombination coefficients, respectively.

\begin{equation}
\alpha_{rad} (A^{i+1}) = \sum_{n=n_{v}} \alpha_{n} (A^{i+1}),
\end{equation}

in which $\alpha_{n} (A^{i+1})$ are the radiative recombination coefficients to level $n$, from the principal quantum number of the ground state of valence $n_{v}$.

  The computed radiative recombination coefficients are fitted to the expression:

\begin{equation}
\alpha_{rad}(T_{e}) = A_{rad} \left( \frac{T_{e}}{10^{4}} \right)^{- \eta},
\end{equation}

where $A_{rad}$ and $\eta$ are the fitting parameters \citep{c27}.

  As far as the dielectronic recombination reactions is concerned, the Burgess's general expression \citep{c28} is extended as \citep{c27}:

\begin{equation}
\alpha_{di}(T_{e}) = A_{di} T_{e}^{-3/2} e^{T_{0}/T_{e}} \left( 1 + B_{di} e^{T_{1}/T_{e}} \right)^{- \eta},
\end{equation}

where all fitting parameters for expressions are reported in appropriate tables.

\subsection{Collisions for thermonuclear reactions in the gas phase}

  As far as the nuclear reactive flows are concerned, the entire landscape of nuclear reactions is depicted either as nuclear decays or as the collision of two nuclear reactants. Using the Maxwell-Boltzmann statistical distribution in the particle velocity, referring to the energy $E$ as integration variable in the centre of momentum frame, as a first approximation the collisional rate coefficient in cm$^{3}$ s$^{-1}$ is \citep{c27}:

\begin{equation}
C_{reac, AB}^{nucl}(T) = < \sigma v > \simeq \frac{(8/\pi)^{1/2}}{m^{1/2} (K_{B} T)^{3/2}} \int \sigma E e^{- \frac{E}{K_{B} T}} dE.
\end{equation}

  The availability of experimental data affects the ability to calculate the integral (54). However, especially for the low temperature ranges ($T < 10^{7}$ K), the statistical extrapolation of data of experimental nonresonant cross sections is customary. In so far as experimental data are available,

\begin{eqnarray}
C_{reac, AB}^{nucl}(T) & \simeq & \Sigma(E^{1/2}) = \Sigma(0) \nonumber \\ & & \left[ 1 + \frac{\dot{\Sigma}(0)}{\Sigma(0)} E^{1/2} + \frac{1}{2} \frac{\ddot{\Sigma}(0)}{\Sigma(0)} E \right],
\end{eqnarray}

for nonresonant neutron cross sections, where the 1st term of the McLaurin series in the $E^{1/2} \propto K_{B} T \propto v$ refers to the 2nd quantum number $l = 0$ s-wave term $\Sigma(0) \propto \pi \lambda^{2} \Gamma_{n}(0) v$. $\Sigma(0)$ is $v$ independent because the De Broglie wavelength $\lambda \propto v^{-1}$ and the partial width for neutron emission $\Gamma_{n}(l) \propto (E E_{r})^{-1/2} \propto m v \hbar^{-1} r_{int}$, where $h$ is the Planck constant, $r_{int}$ is the interaction radius and $E_{r}$ is the centrifugal energy barrier. In the above expression, $\Sigma(0)$, $\dot{\Sigma(0)}$ and $\ddot{\Sigma(0)}$ are empirical constants and differentiations are made with respect to $E^{1/2}$.

  Instead, for charged reactants, nonresonant cross sections are computed by the McLaurin series in the energy $E$: $\Sigma(E)$ as

\begin{equation}
\sigma \simeq \frac{\Sigma(E)}{E} e^{- (E_{G}/E)^{1/2}},
\end{equation}

where

\begin{equation}
\Sigma(E) = \Sigma(0) \left[ 1 + \frac{\dot{\Sigma}(0)}{\Sigma(0)} E + \frac{1}{2} \frac{\ddot{\Sigma}(0)}{\Sigma(0)} E^{2}\right],
\end{equation}

within the integral for $C_{reac, AB}^{nucl}(T)$ as:

\begin{equation}
C_{reac, AB}^{nucl}(T) \simeq \frac{(8/\pi)^{1/2}}{m^{1/2} (K_{B} T)^{3/2}} \int \Sigma(E) e^{- \left( \frac{E_{G}}{E} \right)^{1/2} - \frac{E}{K_{B} T}} dE,
\end{equation}

being $E_{G} = (2 \pi \phi Z_{A} Z_{B})^{2} m c^{2}/2$ the Gamow energy, where $Z_{A}$ and $Z_{B}$ are the atomic numbers of reactants, $\phi$ is the fine structure constant and $c$ is the speed of light. In so doing, the charged particle cross sections (eq.58) are factorized three terms:

\begin{itemize}

\item in a slowly varying energy factor $\Sigma(E)$, decreasing fairly rapidly with increasing $l$ for all partial waves, 

\item in a $E^{-1}$ term coming from the $\pi \lambda^{2}$ and

\item in the Gamow exponential.

\end{itemize}

  In the expression (59) the $\Sigma(0)$, $\dot{\Sigma}(0)$ and $\ddot{\Sigma}(0)$ are empirical constants, but differentiations are made with respect to $E$.

  Whenever a single nuclear resonance occurs, the classical Breight-Wigner formula is largely used, where

\begin{equation}
\sigma = \frac{\pi \hbar^{2}}{2 m E} \frac{\omega_{r} \Gamma_{1} \Gamma_{2}}{(E - E_{r})^{2} + \Gamma^{2}} dE,
\end{equation}

being $E_{r}$ the resonant energy in the centre of momentum system for reactants $A$ and $B$, $\omega_{r} = (1 + \delta_{k_{A} k_{B}})g_{r} (g_{A} g_{B})^{-1}$, where $g_{r}$, $g_{A}$ and $g_{B}$ are the spins $2J + 1$ of the resonant state and reactants $A$ and $B$, respectively. $\Gamma_{1}$ is the partial width for the decay of the resonant state by the emission of the same $A + B$ reactants, while $\Gamma_{2}$ is the partial width for emission of $B + C$ nuclei and $\Gamma = \sum_{k = 1}^{N} \Gamma_{k}$ over all nuclei involved in the reaction.

  In conclusion, as far as the nuclear reactions are concerned, formulae (56 - 61), together with experimental data and with low energy extrapolations, give largely used data tables and approximation fitting formulae \citep{c29,c30,c31,c32,c33}.

\subsection{The setting of the collisional reaction rate coefficients}

  As for the chemical reactions here considered,

\begin{equation}
C_{reac, AB}^{chem}(T) = C_{reac, AB}(T) + \alpha_{rad}(T_{e}) + \alpha_{di}(T_{e}),
\end{equation}

by the summation of expressions (48, 50, 52, 54 and 55), while

\begin{equation}
C_{AB}(T) = C_{reac, AB}^{chem}(T) + C_{reac, AB}^{nucl}(T),
\end{equation}

also including the thermonuclear contribution.

  As far as the thermonuclear reactions are concerned, both nuclear reaction tables, and a lot of fitting formulae exist \citep{c29,c30,c31,c32,c33}. We choose to adopt the fitting formulae reported in these papers, avoiding any interpolation-extrapolation uncertainties due to the arbitrariety of the statistical fitting.

  We would like to include also the ion-electron recombination processes in the code. However, since a general fitting formulation, or some general tables of coefficient for ion-electron recombination dissociations

\begin{equation}
(AB)^{i+j+1} + e^{-} \rightarrow A^{i} + B^{j}
\end{equation}

are still unpublished, we cannot consider at the moment also these reactions. Moreover, although the general formulations for the rate coefficients could also consider molecules-photons interactions, since photon number densities could be obtained knowing their fluxes, we have to postpone such themes because the code we use is free of any component dealing with radiation.

  For a general bimolecular collisional reaction $r$ as: $s_{Ar} A + s_{Br} B \rightarrow s_{Cr} C + s_{Dr} D + ..... $, or as $s_{(AB)r} (AB) \rightarrow s_{Ar} A + s_{Br} B$ formulation (46) can synthetically be rewritten as:

\begin{eqnarray}
\dot{n}_{k}|_{r} & = & C_{k_{A} k_{B}}(T) n_{k_{A}} n_{k_{B}} \frac{s_{kr}}{s_{k_{A}r} s_{k_{B}r}} \nonumber \hspace {0.75cm} \mbox{order $2$ reaction} \\
\mbox{and} & & \\
\dot{n}_{k}|_{r} & = & C_{k_{(AB)}}(T) n_{k_{(AB)}} \frac{s_{kr}}{|s_{k_{(AB)}r}|} \nonumber \hspace{1.1cm} \mbox{order $1$ reaction}
\end{eqnarray}

where the nonzero stoichiometric coefficients $s_{k_{A}r}$, $s_{k_{B}r}$ and $s_{k_{(AB)}r}$ for reaction $r$, for reactants $A$, $B$ and $(AB)$ are assumed negative, while those relative to the reaction products are assumed positive. The general stoichiometric coefficient $s_{kr}$ coincides with that relative to the chemical species $k$ (either a reactant, or a product). Since our code works using the mass density, the mass density rate counterpart of (eq. 65) is:

\begin{eqnarray}
\dot{\rho}_{k}|_{r} & = & C_{k_{A} k_{B}}(T) N_{A} \rho_{k_{A}} \rho_{k_{B}} \frac{\mu_{k}}{\mu_{k_{A}} \mu_{k_{B}}} \frac{s_{kr}}{s_{k_{A}r} s_{k_{B}r}} \nonumber \\
& & \mbox{order $2$ reaction and} \\
\dot{\rho}_{k}|_{r} & = & C_{k_{(AB)}}(T) \rho_{k_{(AB)}} \frac{\mu_{k}}{\mu_{k_{(AB)}}} \frac{s_{kr}}{|s_{k_{(AB)}r}|} \nonumber \\
& & \mbox{order $1$ reaction} \nonumber
\end{eqnarray}

because $\rho_{k} = \mu_{k} m_{H} n_{k} = \mu_{k} n_{k} N_{A}^{-1}$. Formulations (65, 66) are a general extension well working for bimolecular reactions, even if the stoichiometric coefficients are greater than $1$. Whenever $A = B$, the $A + A$ bimolecular (binuclear) collision should be thought as $A + A \equiv 2A$, so that $s_{k_{AA}r} = -2$. In so doing, even the $\delta_{k_{A} k_{B}}$ (eq. 45) is implicitly considered.

  It is important to specify that $C_{AB}(T)$ for order 1 reactions, corresponding to the decay/dissociation constant $D_{k_{(AB)}}$ as: $C_{AB} \equiv C_{(AB)} = D_{k_{(AB)}}$ (\S 3,2), equals $\tau_{k_{(AB)}}^{-1}$, where $\tau_{k_{(AB)}} = D_{k_{(AB)}}^{-1}$ is the decay/dissociation time for reactant $(AB)$. Of course, dealing with an order 1 reaction, we have to decide which formulation we intend to use within the code, that is either a typical decay law or an Arrhenius-like formulation.

\section{The reaction rate algorithm: its building-up and its placement into the fluid dynamics code}
\label{}

\subsection{The initial input of chemical data}

\begin{figure}
\begin{center}
\scalebox{0.2}{\includegraphics[clip=true]{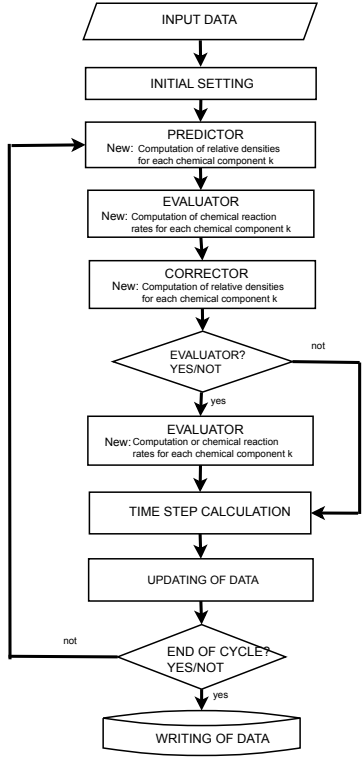}}
\caption{General flow chart of the PEC-PECE integration scheme of eqs. (1-5). The placement of the code sections relative to the reactive flows are schematically shown, without any deepening, at the moment.}
\end{center}
\end{figure}

  The first step in the code implementation is totally arbitrary since several parameters need to be introduced as far as the storing of chemical data within the flow dynamics code is concerned. Each chemical species and each reaction must be labeled by arbitrary identifier numbers. As an example, considering only the two bimolecular chemical reactions $O_{2} + C \rightarrow O + CO$ and $O + CO \rightarrow O_{2} + C$, even including another chemical species (e.g. $CO_{2}$), which is arbitrarily inactive in these two reaction processes, we need to use the reaction label identifier $r = 1, 2$ for the two above reactions, and the chemical label identifier $k = 1,2,...,5$ for $CO_{2}$, $C$, $CO$, $O_{2}$ and $O$, respectively. This arbitrary ordering, of course, could be different, but should be assigned just once from the beginning, taking into account that also the charge status, the isotopic composition, isomers of the same chemical species, and also electrons, involve different identifiers.

  The molecular reaction rate coefficients $\alpha$, $\beta$ and $\theta$ (eq. 48), $A_{rad}$ (eq. 54), $A_{di}$, $B_{di}$, $T_{0}$ and $T_{1}$ (eq. 55) for the computation of $C_{AB}(T)$ (eq. 63) in the code should be set as $\{ K, K, K, R \}$ 4D arrays (that is $\{ 5, 5, 5, 2 \}$ 4D arrays for the above example), whose nonzero elements are only those relative to the appropriate $(k, r)$ pair. The 1st index $k$ in the $\{ k, k_{A}, k_{B}, r \}$ quartet of the above coefficients identifies the chemical label of any molecule, atom, ion or nucleus, etc. included in the list of chemical species, while $k_{A}$ and $k_{B}$ identify the chemical labels of the two reactants. These arrays are symmetric in the index exchange ($k_{A}$, $k_{B}$) to ($k_{B}$, $k_{A}$). The index $r$ in the above $\{ k, k_{A}, k_{B}, r \}$ quartet is related to the reaction label identifier, according to an assumed reaction ordering list.

  Another befitting label identifier $i_{d}$, set as a $\{ R \}$ 1D array, is also used. Whenever we deal with molecular reactions $i_{d} = 0$. From $i_{d} = 1$ onwards, we deal with nuclear reactions. The positive value for $i_{d}$ addresses to the analytical formulation for the chemical nuclear reaction rate coefficient, according to an established order in which nuclear reaction fits are listed in the code as it is made in \citep{c29,c30,c31,c32}. In so doing, $i_{d} = 1$ addresses to the 1st nuclear reaction, $i_{d} = 2$ addresses to the 2nd nuclear reaction and so on. This is unavoidable, since the analytical fitting formulations for nuclear reaction rate coefficients are different from each other, while all chemical molecular reactions rate coefficient formulations refer to the same algebraic formulation (eqs. 48, 54, 55, 62), whose parameters are also largely found in \citep{c2,c6,c7,c34} for molecular combustion thermochemical reactions, as well as in chemical combustion monographs cited at the beginning of this paper.

  Chemical data in the form of $\{ K \}$ 1D arrays for each chemical species, also need to be introduced as far as the initial abundance $X_{k,t=0}$, the molecular weight $\mu_{k}$, the standard enthalpy of formation $\Delta H^{0}_{f, k}$, the molecular-nuclear mass (in uma) $q_{k}$, the number of particles in each molecular compound $N_{p, k}$, the number of molecular bounds $n_{bou, k}$, are concerned. $n_{bou, k}$ values for each molecule are useful for evaluating the $\gamma = c_{p} c_{V}^{-1}$ values in the state equation. Despite still not used in the following, even the degree of ionization $i_{ion, k}$ should also be arranged as a $\{ K \}$ 1D array. Moreover, even the constants of nuclear decay (or the coefficient of molecular dissociations, whenever it is constant) $D_{k, r}$, as well as the stoichiometric coefficients $s_{k, r}$, need to be included as $\{ K, R \}$ 2D arrays.

  Therefore, the initial data setting should include all fluid dynamics usual data. If working in SPH, the initial data setting for each $i$th particle are: $\rho_{i}$, $\dot{\rho}_{i}$, $T_{i}$, $\mathbf{v}_{i}$, $\dot{\mathbf{v}}_{i}$, $\epsilon_{i}$, $\dot{\epsilon}_{i}$, together with the abundance of chemicals $X_{i, k}$ and their mass density $\rho_{i, k}$, as well as $\alpha_{(k, k_{A}, k_{B}, r)}$, $\beta_{(k, k_{A}, k_{B}, r)}$, $\theta_{(k, k_{A}, k_{B}, r)}$, $A_{rad-(k, k_{A}, k_{B}, r)}$, $\eta_{(k, k_{A}, k_{B}, r)}$, $A_{di-(k, k_{A}, k_{B}, r)}$, $B_{di-(k, k_{A}, k_{B}, r)}$, $T_{0-(k, k_{A}, k_{B}, r)}$, $T_{1-(k, k_{A}, k_{B}, r)}$, $i_{d, (k, k_{A}, k_{B}, r)}$, $\mu_{(k, r)}$, $\Delta H^{0}_{f, k}$, $\varepsilon_{dec, k}$, $\varepsilon_{diss, k}$, $\varepsilon_{a, k}$, $Z_{k}$, $N_{p, k}$, $n_{bou, k}$, $i_{ion, k}$, as above discussed, where $\varepsilon_{a, k} = E_{a,k} \mu^{-1}_{k} N_{A}$.

\subsection{General scheme of the code}

  Figure 2 shows the flow chart of the general scheme of a PEC-PECE code in its main junctions. If working in the SPH formalism, since $\dot{\rho}_{ik}|_{r}$, $\dot{\rho}_{ik} = \sum_{r} \dot{\rho}_{ik}|_{r}$ and $\dot{\rho}_{i} = \sum_{k, r} \dot{\rho}_{ik}|_{r}$, must be calculated, together with the flow spatial derivatives for eqs. (1, 3) (e.g. eqs. 12, 14 in SPH), their computations are linearly queued in the Evaluator section of the code right after the code sections dealing with $\dot{\rho}_{i}$. Besides, the same logical setting is also necessary for the computation of $\dot{\epsilon}_{ik}|_{r}$ and $\dot{\epsilon}_{ik}$ just after the SPH computation of $\dot{\epsilon}_{i}$. $\dot{\epsilon}_{ik}|_{r} = - (\varepsilon_{k}/\rho_{i}) \dot{\rho}_{ik}|_{r}$ and $\dot{\epsilon}_{ik} = - \varepsilon_{k} \rho_{i}^{-1} \sum_{r} \dot{\rho}_{ik}|_{r}$, where $\varepsilon_{k}$ corresponds either to $\Delta H^{0}_{f, k}$, or to $\varepsilon_{diss, k}$, or to $q_{k}$ in units of energy per unit mass. As far as the integration of the entire continuity and energy equations are concerned, their usual setting in the Predictor and in the Corrector sections are unchanged. It is important to note that the adoption of eq. (66), gives exactly $\sum_{k=1, r=1}^{K, R} \dot{\rho}_{ik}|_{r} = 0$ for each $i$th particle, as it should be for molecular chemistry, because of the mass conservation. This balance occurs because of the algebraic (positive and negative) stoichiometric coefficients in so far as the molecular weights are written as integer numbers.

\begin{figure}
\begin{center}
\scalebox{0.25}{\includegraphics[clip=true]{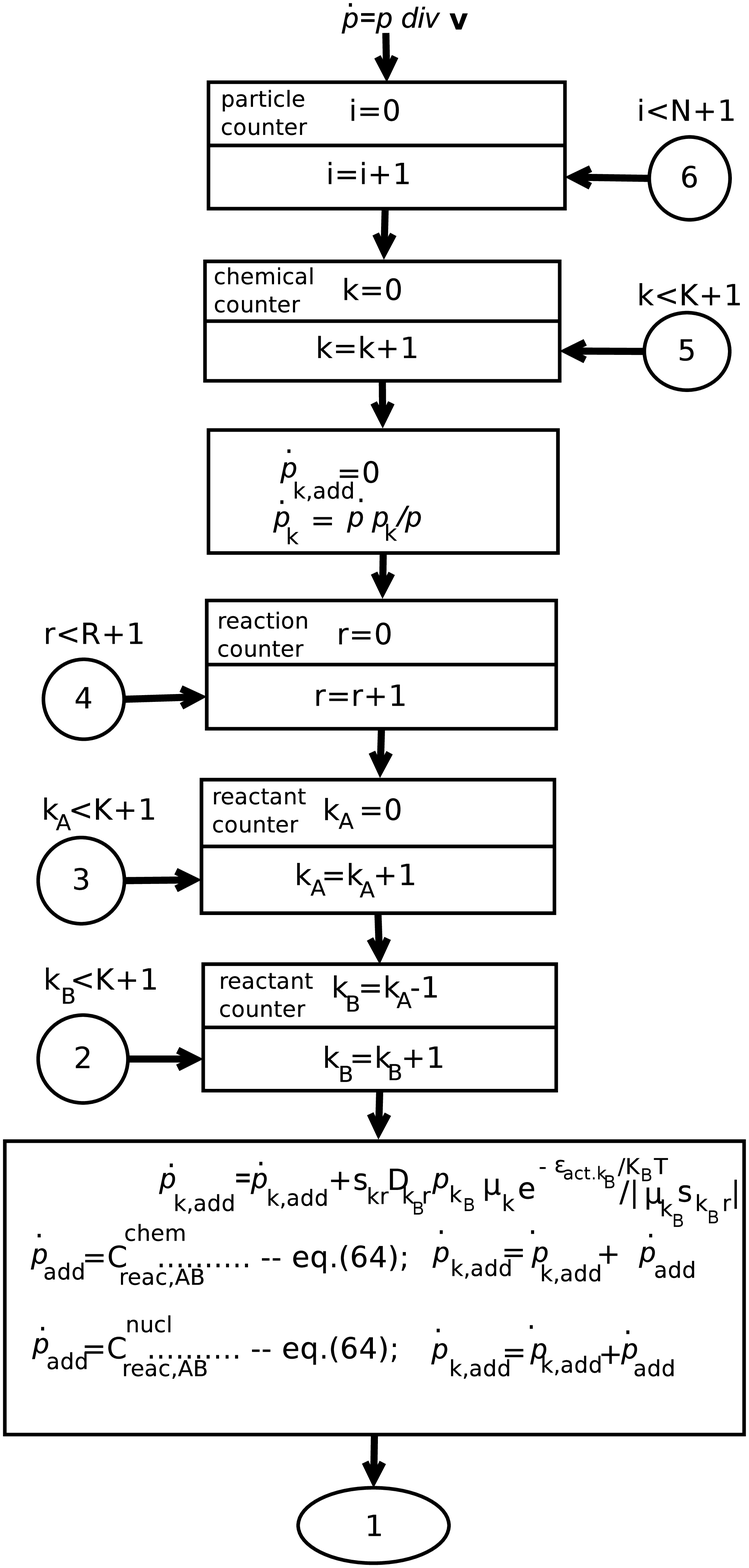}}
\caption{Flow chart relative to the logical scheme for the computation of $\dot{\rho}_{ik}$ in the continuity equation. Subscripts $i$, $k$ and $r$ refer to the $i$th flow or fluid element, to the $k$th chemical component and to the $r$th reaction, respectively. Notice that $D_{kr}$ corresponds to $C_{k_{(AB)}}$ for order 1 reactions, and that subscript $k_{B}$ corresponds to subscript $k_{(AB)}$ of formulation (66). $\dot{\rho}_{add}$ is time by time computed, while $\dot{\rho}_{k,add}$ is algebraically cumulating $\dot{\rho}$ for each chemical species $k$ for each reaction $r$.}
\end{center}
\end{figure}

\begin{figure}
\begin{center}
\scalebox{0.2}{\includegraphics[clip=true]{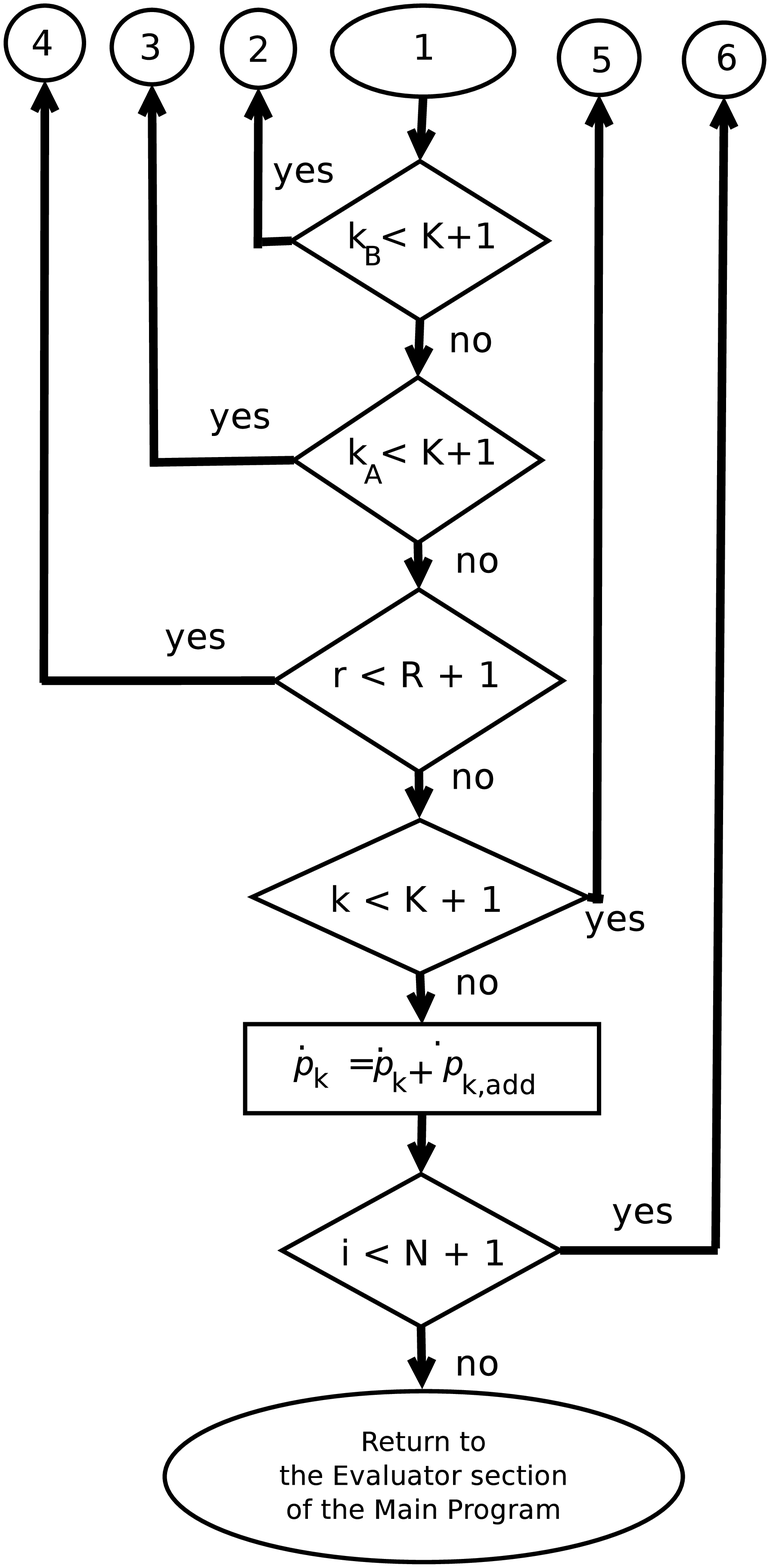}}
\caption{Continuation of Fig. 3.}
\end{center}
\end{figure}

  Figures 3 and 4 show the logical flow charts relative to the calculation of the rates of variations of mass densities for the $i$th particle, for each $k$th chemical species $\dot{\rho}_{k}$ (omitting the subscript $i$ because the logical flow chart scheme is not strictly attributed to SPH schemes). As it is shown on the top of Fig. 3, the unreactive flow mass density variations $\dot{\rho}$ are delivered as input data to compute that part of $\dot{\rho}_{k}$ relative to the local flow variations. Besides, since subscripts $AB$ and $k_{AB}$ are explicitly used in the mathematical formulations for order 1 reactions, the $AB$ notation is omitted both within the logical algorithm and within the code.

  The module, dealing with the new terms, relative to the reactive chemistry includes the calculations giving either the time derivative of the chemical abundances, or, as we prefer, the time derivative of that part of the mass density, dealing with the chemical reactions. In fact, since

\begin{equation}
n_{k} = X_{k} n,
\end{equation}

\begin{equation}
\dot{n} = - n \nabla \cdot \mathbf{v}
\end{equation}

and

\begin{equation}
\dot{n}_{k} = - n_{k} \nabla \cdot \mathbf{v} + \sum_{r} \dot{n}_{k}|_{r},
\end{equation}

that is

\begin{equation}
n^{-1} \dot{n}_{k} = - X_{k} \nabla \cdot \mathbf{v} + n^{-1} \sum_{r} \dot{n}_{k}|_{r},
\end{equation}

so

\begin{equation}
\dot{X}_{k} + X_{k} n^{-1} \dot{n} = -X_{k} \nabla \cdot \mathbf{v} + n^{-1} \sum_{r} \dot{n}_{k}|_{r},
\end{equation}

that (from eq. 68) becomes:

\begin{equation}
\dot{X}_{k} = \sum_{r} \dot{X}_{k}|_{r} = n^{-1} \sum_{r} \dot{n}_{k}|_{r}, \hspace{.2 cm} \mbox{and} \hspace{.2 cm} \dot{X}_{k}|_{r} = n^{-1} \dot{n}_{k}|_{r}.
\end{equation}

Instead, as we prefer, considering

\begin{equation}
\dot{n}_{k} = X_{k} \dot{n} + \dot{X}_{k} n,
\end{equation}

\begin{eqnarray}
\dot{\rho}_{k} & = & - X_{k} \mu_{k} \overline{\mu}^{-1} \rho \nabla \cdot \mathbf{v} + \dot{X}_{k} \mu_{k} \overline{\mu}^{-1} \rho \nonumber \\
& = & - \rho_{k} \nabla \cdot \mathbf{v} + \sum_{r} \dot{X}_{k}|_{r} \mu_{k} \overline{\mu}^{-1} \rho \nonumber \\
& = & - \rho_{k} \nabla \cdot \mathbf{v} + \sum_{r} \dot{\rho}_{k}|_{r},
\end{eqnarray}

being (from eq. 72)

\begin{equation}
\dot{X}_{k}|_{r} = \rho^{-1} \overline{\mu} \mu_{k}^{-1} \dot{\rho}_{k}|_{r}.
\end{equation}

  In addition to one of the two routes (eq. 72 or eq. 74), the standing alone algorithm, dealing with chemistry in the Evaluator section of the code, must also include the corresponding thermal energy power per unit mass given by the chemistry alone.

\begin{figure}
\resizebox{\hsize}{!}{\includegraphics[clip=true]{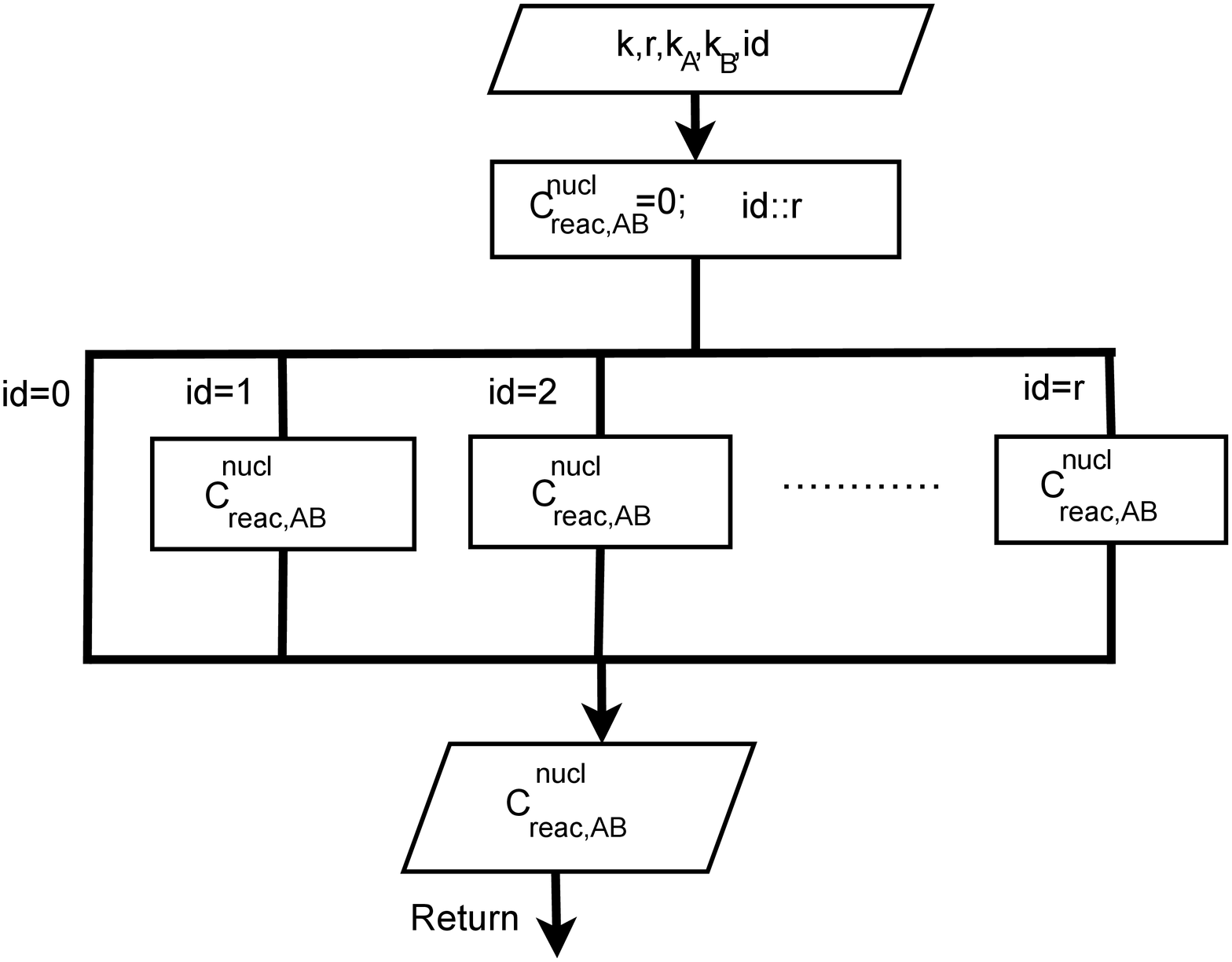}}
\caption{Logical flow chart for the computation of the nuclear reaction rate coefficients. For shortness reasons, the algebraic fitting formulae for each $C_{reac, AB}^{nucl}$ calculation are not reported.}
\end{figure}

  According to our point of view, this means that (eq. 74) $\dot{\rho}_{k}$ is determined by the $\dot{\rho} \rho_{k} \rho^{-1} = - \rho_{k} \nabla \cdot \mathbf{v}$ contribution, that is the local mass density variations only due to the flow dynamics, plus a second term only due to the reaction processes, without any variation of the local total mass density $\rho$, where the chemical abundance and the mean molecular weight variations are involved. Hence, the first term on the right of eq. (74) is exactly what we included in the scheme of Fig. 3, while $\dot{\rho}_{k}|_{r}$ is evaluated in a different fashion through binary collisions processes through eq. (66), because of chemical processes, without any flow variation. As far as the order 1 reactions are concerned, including the molecular chemical dissociations and nuclear decays, the innermost counter $k_{B}$ in the flow chart of Figs. 3 and 4 is also used to represent the $k_{(AB)}$ subscript in eqs. (49, 66).

\begin{figure}
\begin{center}
\scalebox{0.25}{\includegraphics[clip=true]{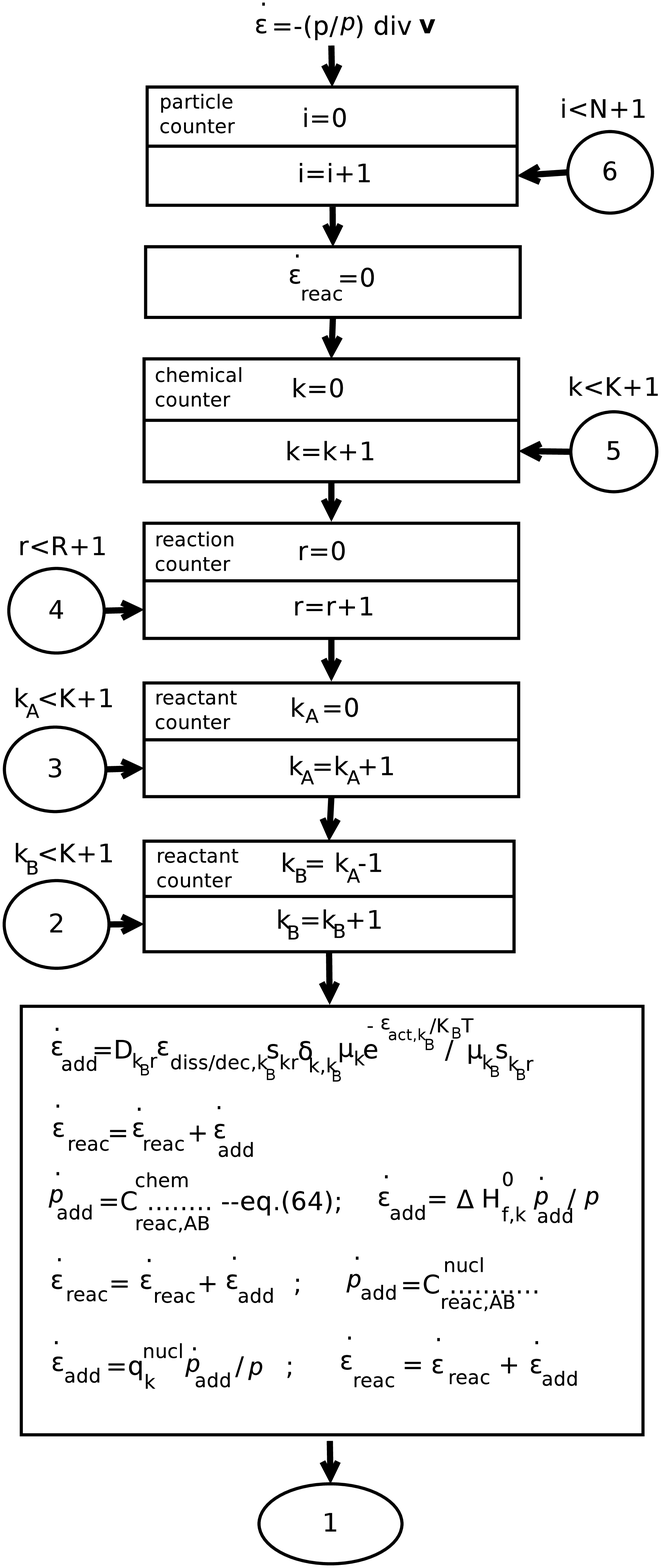}}
\caption{Flow chart relative to the logical scheme for the computation of $\dot{\epsilon}_{i}$ in the energy equation. Subscripts are displaced with the same logical order of previous Figs. 3, 4 and 5. Notice that $D_{kr}$ corresponds to $C_{(AB)}$ for order 1 reactions, and that $\rho_{k_{B}}$ corresponds to $\rho_{k_{(AB)}}$ of formulation (66). $\dot{\epsilon}_{add}$ is computed time by time, while the algebraic cumulating reactive thermal energy power per unit mass is $\dot{\epsilon}_{reac}$.}
\end{center}
\end{figure}

\begin{figure}
\begin{center}
\scalebox{0.2}{\includegraphics[clip=true]{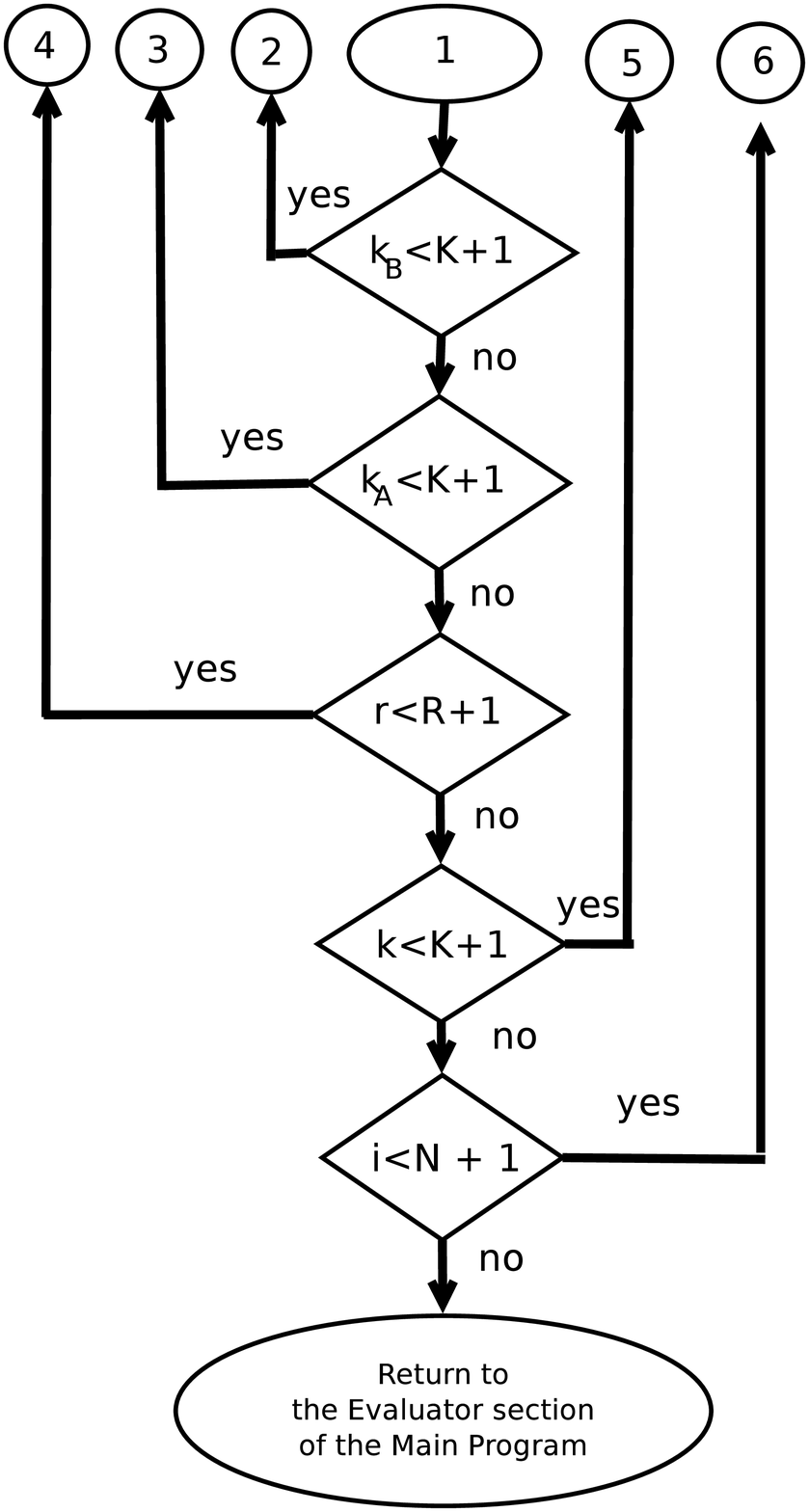}}
\caption{Continuation of Fig. 6.}
\end{center}
\end{figure}

  As it is shown in the flow charts of Figs. 3 and 4 as far as the calculation of $\dot{\rho}_{k}$ is concerned, any nuclear decay or any molecular dissociation are first considered. Of course, the one excludes the other for the same chemical species in the same reaction (or in the same spontaneous nuclear decays, where $\varepsilon_{act, k} \equiv \varepsilon_{act, k_{(AB)}} =0$). Then the calculation relative to binary molecular collisions is considered and finally the calculations relative to any collisional nuclear chemistry are taken into account. This sequence is ordered in nested cycles for each $i$th particle, chemical species, reaction, reactants $A$ and $B$. The logical scheme of nuclear chemistry, being based on appropriate algebraic fitting formulations for each reaction, is structurally very different, as it is shown in Fig. 5, where its logical flow chart is reported. Whenever $i_{d}= 0$, only chemical molecular reactions are considered, so that $C_{reac,AB}^{nucl} = 0$, while $C_{reac, AB}^{chem} > 0$. Instead, whenever $1 \leq i_{d}\leq r$, $C_{reac, AB}^{chem} = 0$, while $C_{reac, AB}^{nucl} > 0$.

  The flow charts dealing with the thermal energy power per unit mass contribution, related to the chemical reactions, are shown in Figs. 6 and 7. Of course these flow charts resemble those shown in Figs. 3 and 4, with the difference that the $\dot{\epsilon}_{reac}$ algebraic accumulation starts just after setting the counter $i$ for the SPH particle, so that it takes into account the entire thermal energy power per unit mass developed by all reactions for all reacting chemicals. In the innermost part of the nested cycles, the contribution coming out from the chemical species breakup (either molecular dissociation or nuclear decay), followed by the bimolecular or binuclear collision contributions, are considered. As it is shown for the mass density chemical rates, the chemical reactive thermal energy power per unit mass contributions $\dot{\epsilon}_{ik}|_{r}$ and $\sum_{k, r} \dot{\epsilon}_{ik}|_{r}$ for each $i$th particle, are linearly queued just after the code section working with $- \rho^{-1} p \nabla \cdot \mathbf{v}$.

  It is important to note that throughout the formulations here expressed for chemical reactions and throughout the flow charts here shown, the index $i$, here dealing with a SPH Free Lagrangian particle identifier, does not concern with any chemical reaction and it is neutral in this aspect. This occurs because we are working in nondiffusive conditions, so that each $i$ particle has its own standing alone chemical evolution in its ($k, r$) indexes. This explain why we claimed that the algorithm we propose could be implemented in a wider scenario also involving other numerical techniques.

\subsection{Time step for reactive flows}

  A time step restriction is always necessary for time dependent calculations in computational fluid dynamics. Currently, such restrictions are needed for mathematical stability reasons in explicit calculations of partial differential equations (PDE).

  For computational flow explicit calculations, the Courant-Friedrichs-Lewy condition \citep{c35,c36} is imposed on those hyperbolic terms representing advection in PDE (spatial derivatives of pressure or velocity), where the given Courant number $C_{CFL} = v_{c} \Delta t_{CFL} (\Delta r)^{-1} \leq 1$ is generally of the order of $0.2 - 0.5$, where $\Delta r$ is the spatial resolution, $v_{c}$ is the maximum value among the local kinematic and the signal transmission velocities within the computational domain, and $\Delta t_{CFL}$ is the Courant-Friedrichs-Lewy time step to be computed.

The Courant-Friedrichs-Lewy condition on the time step progression to solve PDE and ordinary differential equations (ODE) for explicit integration techniques offers a temporal reference where numerical solutions are both stable and convergent with the mathematical solutions.

  For SPH technique, the explicit time limiter is given by:
  
\begin{equation}
\Delta t_{SPH}^{CFL} = C_{CFL} \cdot {\textrm {min}}_{i=1,N} \left[ \frac{h_{i}}{v_{sig,ij}}, |\nabla \cdot \boldmath{v}_{i}|^{-1}, \left( \frac{h_{i}}{|\boldmath{a}_{i}|} \right)^{1/2} \right],
\end{equation}

which includes the Courant-Friedrichs-Lewy time limiter $\Delta t_{CFL}$. $v_{sig,ij}$ is the signal transmission velocity between close particles $i$ and $j$ within the SPH spatial resolution length $h_{i}$ \citep{c9,c11,c12,c37}, also including the sound speed $c_{si}$, while $|\mathbf{a}|_{i}$ is the full acceleration for the $ith$ SPH particle. $C_{CFL}$ is a number of the order of $0.2 - 0.5$.

  Including the reactive processes, the reaction speed for each chemical reaction, for each chemical species needs also to be taken into account, for each particle $i$. This term has to be included in the computation for the explicit time limiter (eq. 76) as:

\begin{eqnarray}
v_{reac,i} & = & h_{i} \cdot {\textrm {max}}_{k=1, K}^{r=1, R} \left[ \frac{\rho_{i}}{\dot{\rho}_{ik}} \right]_{r} \\
\Delta t_{SPH}^{reac} & = & C_{CFL} \cdot {\textrm {min}}_{i=1,N} \Bigg[ \frac{h_{i}}{v_{sig,ij}}, |\nabla \cdot \boldmath{v}_{i}|^{-1}, \left( \frac{h}{|\boldmath{a}_{i}|} \right)^{1/2}, \nonumber \\ & & \frac{h_{i}}{v_{reac,i}} \Bigg],
\end{eqnarray}

where ${\textrm max}_{k=1, K}^{r=1, R}$ shows how the speed of reactions has to be calculated on all reactions for all chemicals for all particles. This ensures that the flow chemical modifications are consistent with the flow dynamics, in so far as $\Delta t_{SPH}^{reac}$ is not too small with respect to $\Delta t_{SPH}^{CFL}$, otherwise some instabilities could occur.

  Notice that eqs. (76, 77, 78) still hold for more general numerical scheme formalisms substituting the SPH spatial resolution length $h_{i}$ with the local resolution length assumed in other schemes or fluid dynamic description. So that the entire discussion of this subsection is not strictly dedicated to the SPH framework.

\section{Gas phase reactions on 2D planar discontinuity flow fronts}
\label{}

  In the following two subsections we turn our attention to the consequence of the inclusion of reactive terms (eqs. 1-5), consisting of the energy contribution of reactions and of the role that reactions play on the mean molecular weight and on the ratio of specific heats $\gamma$ in the EoS. To do this, we perform some 2D applications in the GASPHER framework in the case of planar discontinuity flow fronts. Any discussion on the purely fluid dynamic aspect, dealing with the adoption of a Gaussian extended interpolation Kernel in GASPHER compared with the traditional cubic spline Kernel in the SPH approach, is given in \citep{c10}, where a complete discussion is shown, dealing with a more complete fluid dynamics.

  In this paper, both molecular and nuclear reactions are taken into account, using the same code, where only access data characterize the wished thermochemistry. The number of reactions considered is voluntarily small because, in so doing, the modifications in the planar discontinuity flow profile can be better understood.

  Throughout the numerical simulations, both reactive and their related unreactive flows, the initial conditions: physical, geometrical and chemical, are exactly the same on both sides of the initial discontinuity.

  The initial setting of GASPHER particles is that of equally displaced particles within a $xy$ box, whose lengths are $50 \times 10$, respectively, where particle mutual separation, as well as their spatial resolution length $h$, are equal to $0.05$. In so doing, a medium size resolution of $10^{-3}$, along the $x$ axis of propagation of flow discontinuity, is taken into account. The two extreme edges along the $x$ horizonthal direction are strictly at rest, since the local velocity and acceleration are forced to be zero within the two vertical edge thicknesses, regarding two rows of vertical edge particles for each. Instead, more simply, no acceleration along the $y$ direction is allowed on the other two other horizonthal edges of the box, along the $y$ direction. Thus, such other edge particles can only flow slipping along the $x$ direction, preventing any outflowing perpendicular to the $x$ direction. The initial flow discontinuity is placed at $x = 25$.

  As the system of equations is dimensionless, the normalization value for the spatial distances $x_{\circ}$ is assumed equal to $10^{2}$ Km for those numerical simulations regarding the molecular chemical reactive flows, while $x_{\circ} = 10^{5}$ Km as far as the nuclear chemical reactive flows are concerned. Those are clearly arbitrary values. The initial sound speed on the right side of the initial discontinuity is assumed as the normalization value $c_{s \circ}$ for the velocities. Hence, $t_{\circ} = x_{\circ} c_{s \circ}^{-1}$ is the normalization value for the time. $\rho_{\circ}$ and $\epsilon_{\circ}$, relative to the initial values for the mass density and the thermal energy per unit mass on the right side of the initial discontinuity, are assumed as the normalization values for $\rho$ and $\epsilon$, respectively, being $\rho_{\circ} = 10^{-14}$ g cm$^{-3}$ wherever it is not explicitly written. $v_{x}$ is normalized to the initial sound speed $c_{s \circ}$.

 Throughout the numerical simulations the entire flow is assumed initially at rest ($v_{x} = v_{y} = 0$), while the left-to-right particle mass ratio - as well as the initial mass density ratio - are both imposed to be equal to $10^{2}$.

\subsection{Molecular thermochemistry on 2D planar discontinuity flows: Reactive and nonreactive flow computational tests}

 The molecular reactions we considered, as well as their Arrhenius fitting data \citep{c34}, are shown in Tab. 1.

\begin{table}
\begin{center}
\begin{tabular}{lccc}

reaction & $\alpha$ & $\beta$ & $\theta$ \\
\hline
$O_{2} + C \rightarrow O + CO$ & $1.05 \cdot 10^{-12}$ & $0.5$ & $0$ \\
$O + CO \rightarrow O_{2} + C$ & $1.67 \cdot 10^{-12}$ & $0.5$ & $69300$ \\
$O_{2} + CO \rightarrow O + CO_{2}$ & $4.18 \cdot 10^{-12}$ & $0.5$ & $25000$ \\
$O + CO_{2} \rightarrow O_{2} + CO$ & $2.84 \cdot 10^{-11}$ & $0$ & $28000$ \\
\end{tabular}
\end{center}

\caption{Arrhenius fitting coefficients for the gas phase chemical reactions. $\alpha$ coefficient units are cm$^{3}$ s$^{-1}$ for two bodies reactants.}

\end{table}


  In this first numerical experiment, first we only consider the two reactions: $O_{2} + C \leftrightarrows O + CO$, assuming that the initial abundances of $C$ and $O_{2}$ are equal to $0.5$ on both sides of the initial discontinuity for each particle. The initial temperature on the right side of the initial discontinuity is $4 \cdot 10^{2}$ K. On the left side the initial temperature is $11$ times larger with respect to the right side. These orders of magnitude make it possible the adoption of the state equation of ideal gases:

\begin{equation}
p = \overline{\mu}^{-1} (\gamma - 1) \rho \epsilon,
\end{equation}

because any further term (e.g. terms dealing with the Van der Waals, Redlich-Kwong EoS, etc.) is negligible. Reactions occur on both sides of the discontinuity. Besides, it is relevant to note that for these first two reactions, $\overline{\mu}$  and $\gamma$ do not vary throughout the reactive processes, so that in this first test any role played by $\overline{\mu}$ and $\gamma$ is excluded.

\begin{figure}
\resizebox{\hsize}{!}{\includegraphics[clip=true]{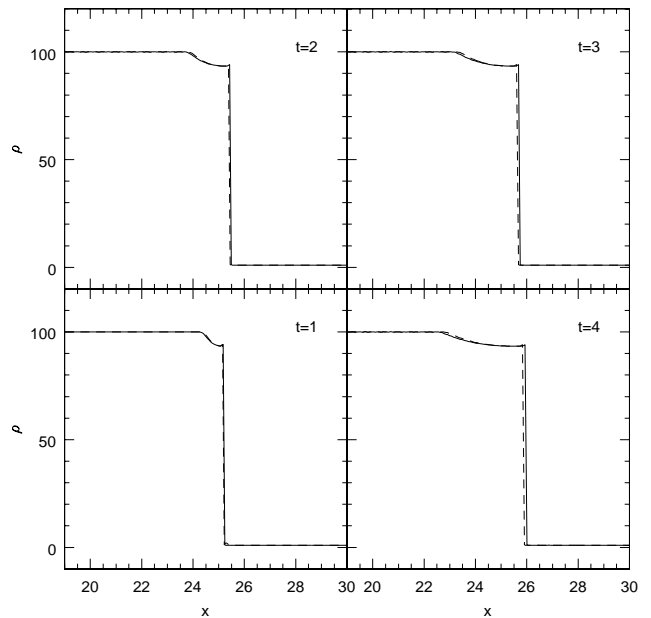}}
\caption{1st model planar discontinuity flow profiles of total mass density $\rho$ at times $t = 1, 2, 3, 4$ for reactive flows where $O_{2} + C \leftrightarrows O + CO$ reactions occur. The unreactive discontinuity flow profile is also shown for comparison in a dashed line profile.}
\end{figure}

\begin{figure}
\resizebox{\hsize}{!}{\includegraphics[clip=true]{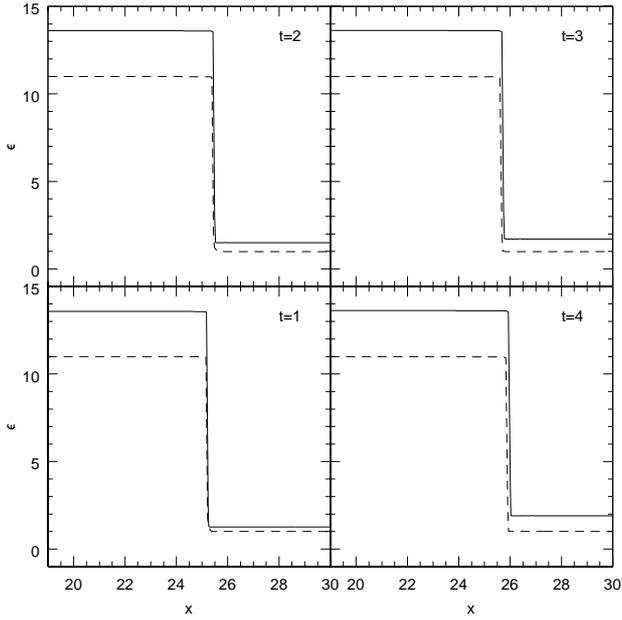}}
\caption{1st model planar discontinuity flow profiles of energy per unit mass $\epsilon$ at times $t = 1, 2, 3, 4$ for reactive flows where $O_{2} + C \leftrightarrows O + CO$ reactions occur. The unreactive discontinuity flow profile is also shown for comparison in a dashed line profile.}
\end{figure}

\begin{figure}
\resizebox{\hsize}{!}{\includegraphics[clip=true]{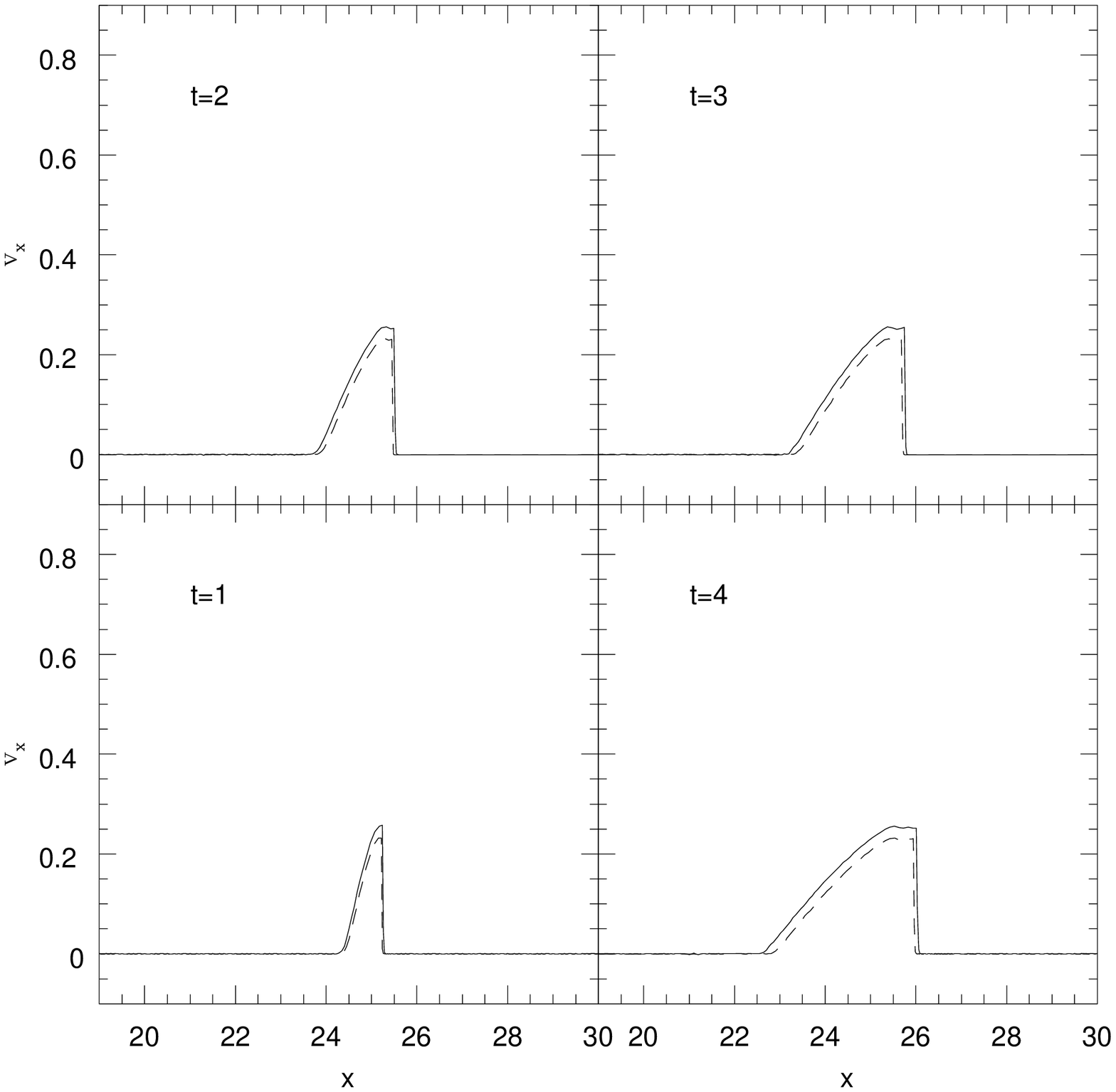}}
\caption{1st model planar discontinuity flow profiles of velocity $v$ at times $t = 1, 2, 3, 4$ for reactive flows where $O_{2} + C \leftrightarrows O + CO$ reactions occur. The unreactive discontinuity flow profile is also shown for comparison in a dashed line profile.}
\end{figure}

\begin{figure}
\resizebox{\hsize}{!}{\includegraphics[clip=true]{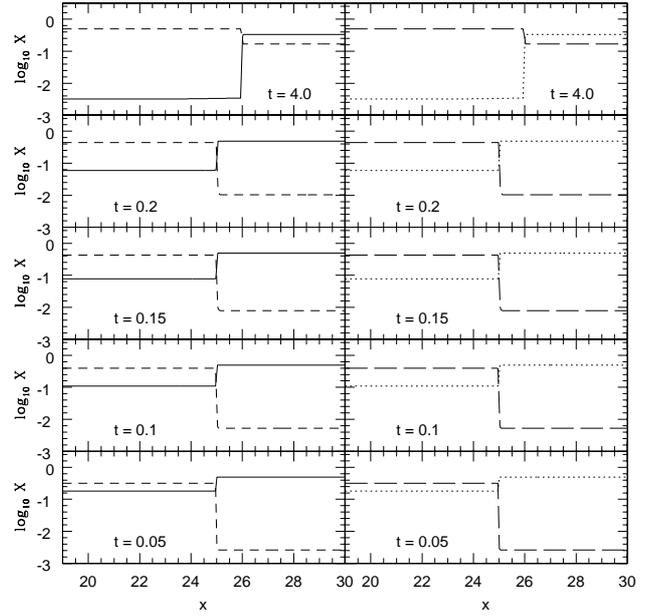}}
\caption{1st model abundance distributions relative to the $O_{2} + C \leftrightarrows O + CO$ reactions. Abundance distributions of $C$ in solid line, and $CO$ in short dashed line on the left panels of the plot at times $t = 0.05, 0.1, 0.15, 0.2, 4$. Abundance distributions of $O_{2}$ in dots, and $O$ in long dashed line on the right panels of the plot at the same instances.}
\end{figure}

 Of the two chemical reactions: $O_{2} + C \rightarrow O + CO$ and $O + CO \rightarrow O_{2} + C$, the first one, exothermic, prevails since the activation temperature $\theta$ for the second one grows stout the role of the second reaction at temperatures $\approx 3 - 6$ times larger than those here found on the hotter left side of the discontinuity flow.

  Figures 8, 9 and 10 show the profile of the discontinuity flow evolution for the mass density $\rho$, the thermal energy per unit mass $\epsilon$ and the velocity $v_{x}$, respectively at times $t = 1, 2, 3, 4$ for both the reactive, as well as for the unreactive flows.

\begin{figure}
\resizebox{\hsize}{!}{\includegraphics[clip=true]{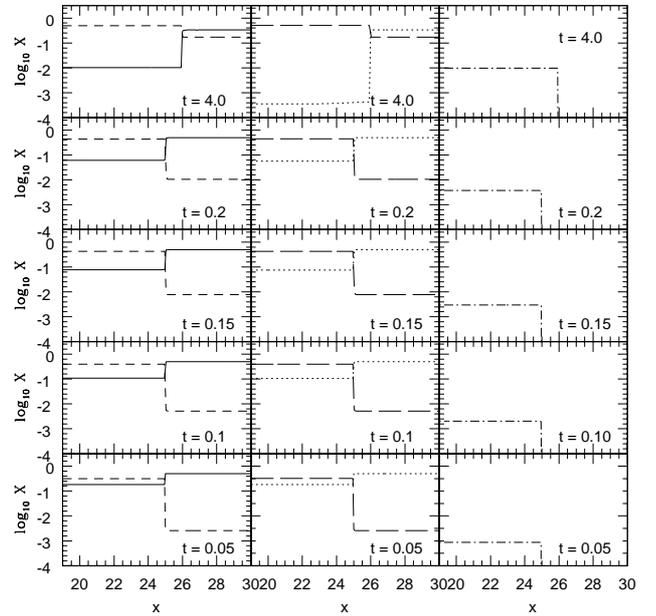}}
\caption{2nd model abundance distributions relative to the $O_{2} + C \leftrightarrows O + CO$ and $CO + O_{2} \leftrightarrows CO_{2} + O$ reactions. Abundance distributions of $C$ in solid line, and $CO$ in short dashed line on the left panels of the plot at times $t = 0.05, 0.1, 0.15, 0.2, 4$. Abundance distributions of $O_{2}$ in dots, and $O$ in long dashed line are also shown on the middle panels of the plot at the same instances. Abundance distribution of $CO_{2}$ in short dashed-dots line are also shown on the right panels of the plot at the same instances.}
\end{figure}

\begin{figure}
\resizebox{\hsize}{!}{\includegraphics[clip=true]{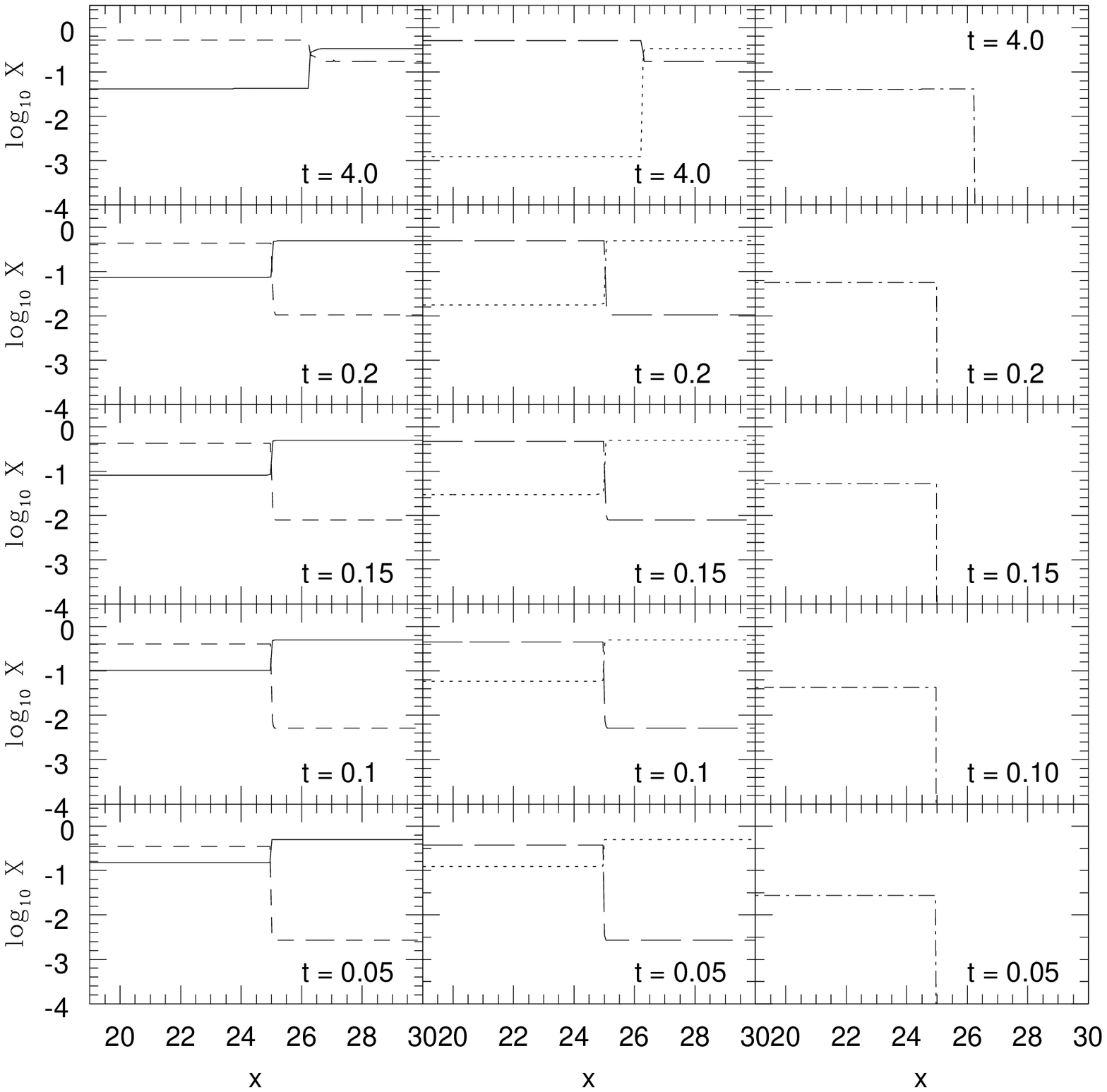}}
\caption{3rd model abundance distributions of $C$, $CO$, $O_{2}$, $O$ and $CO_{2}$, doublingb the initial temperature of the left side discontinuity flow with respect to that relative to Fig. 12.}
\end{figure}

  Figure 8 shows that even in those part of the planar profile, where the temperature ($T \propto \epsilon$) is higher for reactive flows, the total mass density is correctly conserved because of the right balance of the partial mass densities among chemicals. Of course, being the $O_{2} + C \rightarrow O + CO$ an exothermic reaction, the temperature increases throughout, especially in those hotter parts (left side) of the discontinuity flow. A progressive increase of temperature is also visible, over time, on the right side of the discontinuity flow, where chemical reactions are much slower (Fig. 9). Moreover, being the pressure spatial gradients larger at the reactive discontinuity flow position between the two different sides, the reactive flows is progressively spatially a bit more advanced throughout compared with that relative to the unreactive flow.

\begin{figure}
\resizebox{\hsize}{!}{\includegraphics[clip=true]{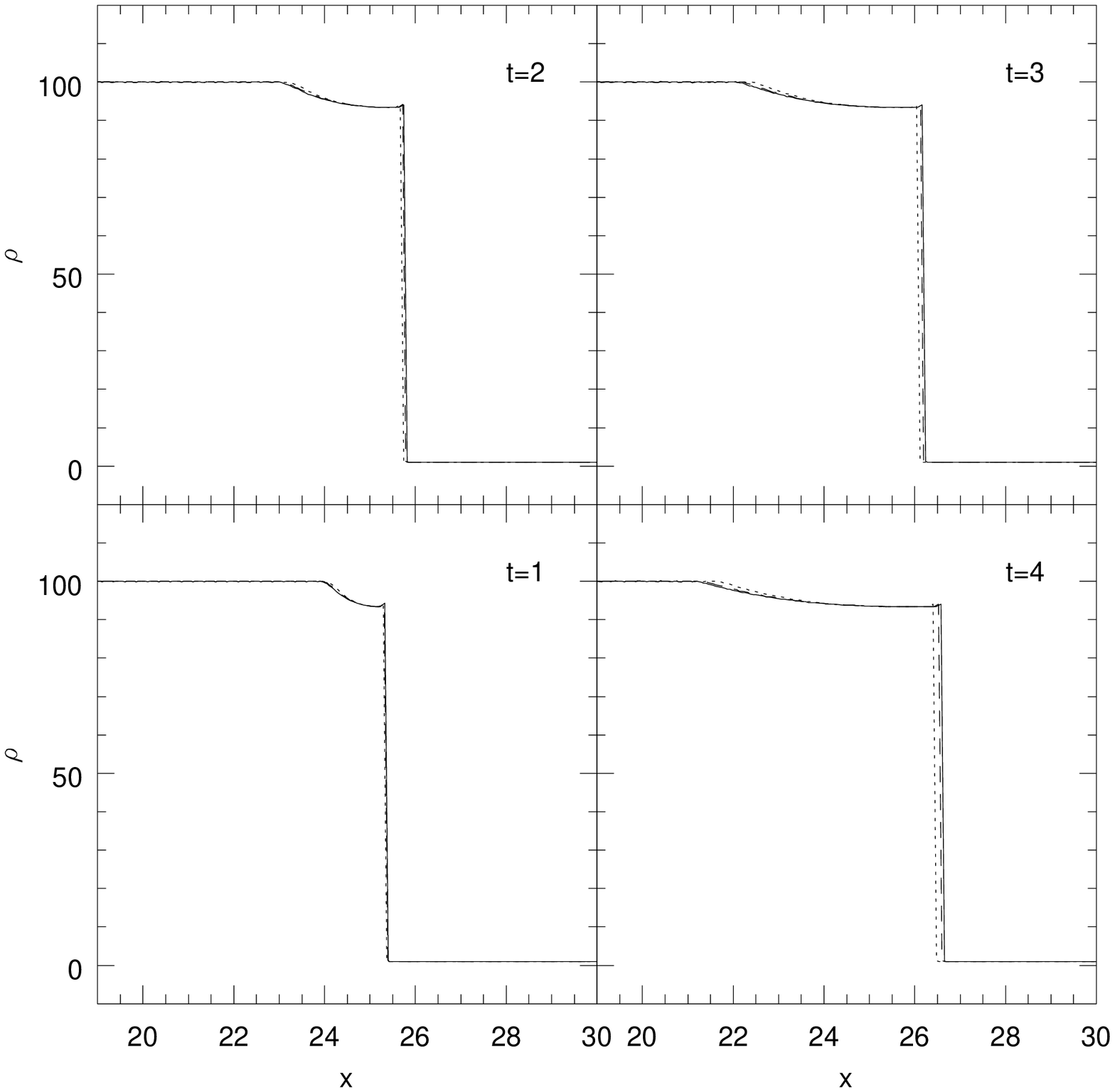}}
\caption{Comparison of planar discontinuity flow profile of the total mass density $\rho$ at times $t = 1, 2, 3, 4$ for reactive flows where $O_{2} + C \leftrightarrows O + CO$ and $CO + O_{2} \leftrightarrows CO_{2} + O$ reactions occur (dots - 4th model). The unreactive discontinuity flow profile is also shown for comparison (dashed line profile), as well as the planar discontinuity flow profile for the reactive flow where only the two $O_{2} + C \leftrightarrows O + CO$ reactions (5th model) are considered.}
\end{figure}

\begin{figure}
\resizebox{\hsize}{!}{\includegraphics[clip=true]{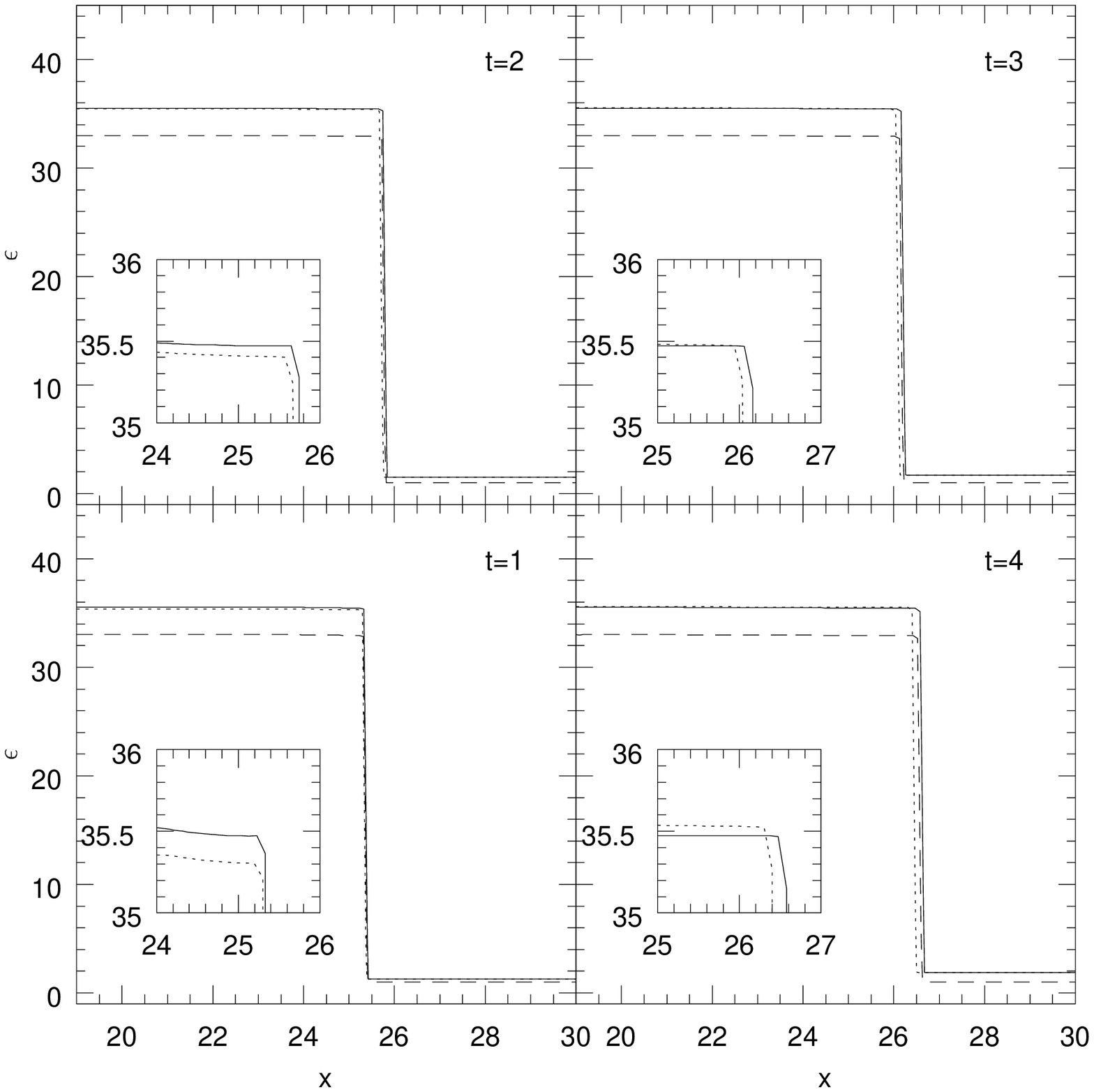}}
\caption{Comparison of planar discontinuity flow profiles of the thermal energy per unit mass $\epsilon$ at times $t = 1, 2, 3, 4$ for reactive flows where $O_{2} + C \leftrightarrows O + CO$ and $CO + O_{2} \leftrightarrows CO_{2} + O$ reactions occur (dots - 4th model). The unreactive discontinuity flow profile is also shown for comparison (dashed line profile), as well as the planar discontinuity flow profile for the reactive flow where only the two $O_{2} + C \leftrightarrows O + CO$ reactions (5th model) are considered. For each panel, an enlargement is also shown, enhancing the profile differences.}
\end{figure}

\begin{figure}
\resizebox{\hsize}{!}{\includegraphics[clip=true]{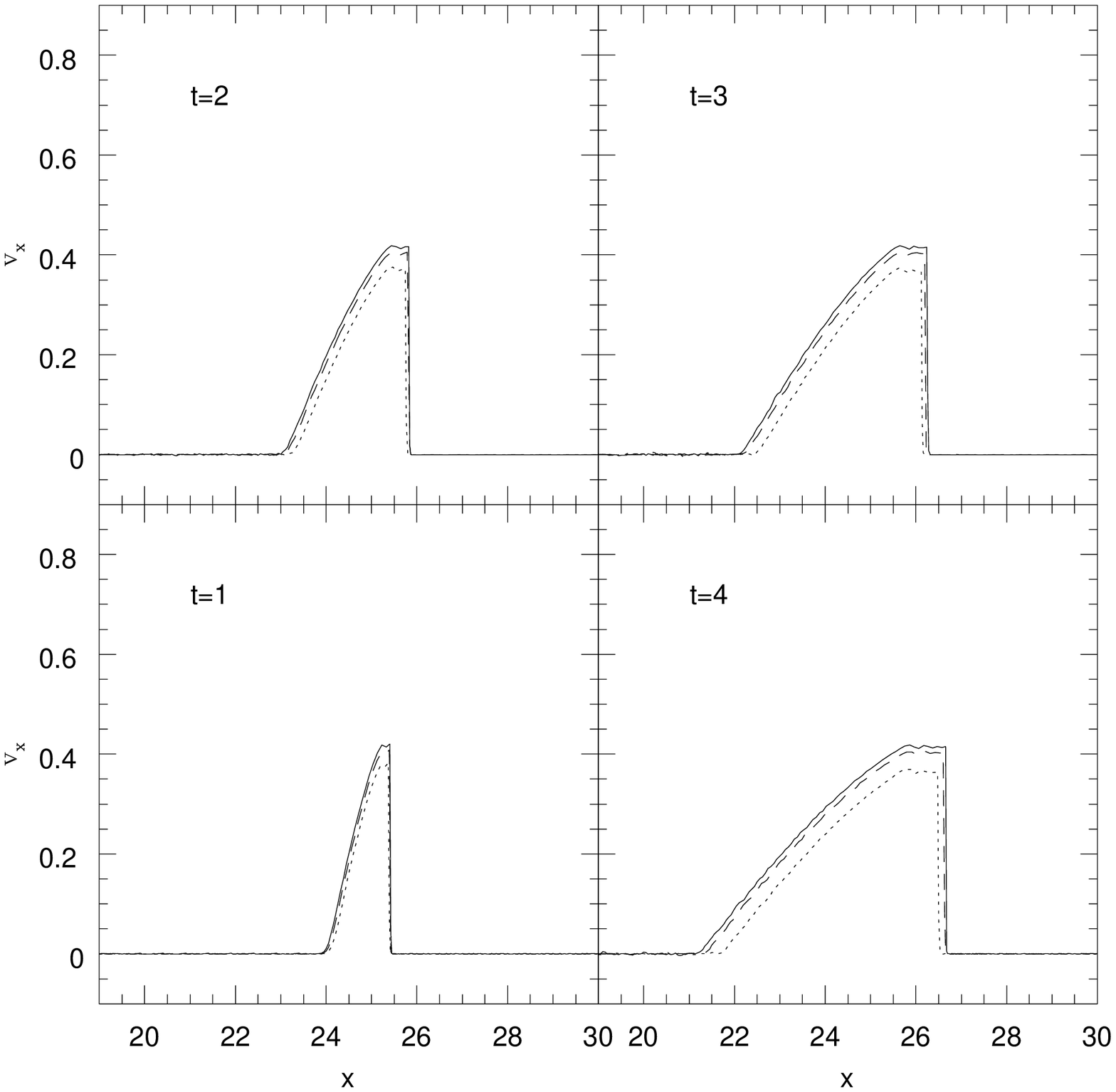}}
\caption{Comparison of planar discontinuity flow profile of the velocity $v_{x}$ at times $t = 1, 2, 3, 4$ for reactive flows where $O_{2} + C \leftrightarrows O + CO$ and $CO + O_{2} \leftrightarrows CO_{2} + O$ reactions occur (dots - 4th model). The unreactive discontinuity flow profile is also shown for comparison (dashed line profile), as well as the planar discontinuity flow profile for the reactive flow where only the two $O_{2} + C \leftrightarrows O + CO$ reactions (5th model) are considered.}
\end{figure}

  This numerical test clearly shows that the structure, the dynamics and the evolution of reactive flows can substantially differ from those relative to unreactive. Of course, in the case of endothermic chemical reactions, the entire argument and conclusions should be seen on the opposite side as a consequence of the local flow reactive cooling.

  The increase of the thermal energy per unit mass on the hotter left side of the reactive discontinuity flow (Fig. 9) is $\approx 2.7$ since $t = 1$ onwards, corresponding to a temperature increase $\Delta T \approx 1.3 \cdot 10^{3} - 1.4 \cdot 10^{3}$ K. The corresponding increase in the thermal energy per unit mass can be obtained considering that $\epsilon_{\circ} = K_{B} N_{A} (\gamma - 1)^{-1} T_{\circ} \approx 6 \cdot 10^{12}$ erg g$^{-1}$. Each chemical reaction $O_{2} + C \rightarrow O + CO$ involves a thermal energy power per unit mass $\upsilon \approx \rho^{-1} \sum_{k} \varepsilon_{k} \Delta \rho_{k}/ \Delta t$. $\varepsilon_{k} = \Delta H^{0}_{f, k} \cdot 10^{10} \mu^{-1}_{k}$ in erg g$^{-1}$, where $\Delta H^{0}_{f, k}$ is usually expressed as Kj mol$^{-1}$. Since at time $\Delta t \approx 10^{-2}$ the most of the combustion process is done, but the entire chemical process is still in progress, a ratio of $\upsilon \Delta t^{-1} \epsilon^{-1}_{\circ} \approx 1.3 \cdot 10^{11}/(10^{-2} \cdot 6 \cdot 10^{12}) \approx 2.2$, that is a crude estimation, but it is not anomalous in comparison with $2.7$. 

  An abundance fraction of the initial reactants, up to some percentages, still exists within $0 < t \leq 0.2$, as it is found in Fig. 11, showing the abundance distributions of $C$, $CO$, $O_{2}$ and $O$ at times $t = 0.05, 0.1, 0.15, 0.2, 4$. On the hotter left side of the flow discontinuity, for each panel plot, the fast abundance distribution evolution is clearly visible up to its saturation-equilibrium. The correctness of results relies on the fact that for times $t\geq 1$ on the left hotter side of the discontinuity flow front, where $\rho$ still holds its initial values, the $\epsilon$ profile is still flat in spite of the reactive heating, and the thermodynamics of reactions is statistically steady, an equilibrium is gradually accomplished. There, the two reaction rates equal with each other, as it is recorded from the molecular abundance distributions from $t = 1$ onwards, where:

\begin{equation}
C_{O_{2} + C \rightarrow O + CO} n_{O_{2}} n_{C} = C_{O + CO \rightarrow O_{2} + C} n_{O} n_{CO},
\end{equation}

being

\begin{equation}
\frac{C_{O_{2} + C \rightarrow O + CO}}{C_{O + CO \rightarrow O_{2} + C}} = \frac{X_{O} X_{CO}}{X_{O_{2}} X_{C}} \approx 2 \cdot 10^{5}.
\end{equation}

  Reversing the initial chemical abundances, assuming $0.5$ as the initial abundance for both $CO$ and $O$ on both sides of the initial discontinuity, there are not relevant differences between the reactive and the unreactive discontinuity flow profiles since the combustion reaction rates for the $CO + O \rightarrow C + O_{2}$ are very small for such initial temperature order of magnitude. Therefore, results dealing with this modelling are not shown.

  Figures 12 and 13 show results for the 2nd and 3rd models as far as the abundance distributions of $C$, $CO$, $O_{2}$, $O$ and $CO_{2}$ are concerned, considering all four reactions $O_{2} + C \leftrightarrows O + CO$ and $CO + O_{2} \leftrightarrows CO_{2} + O$, adopting the same initial conditions (2nd model - Fig. 12), but doubling the initial temperature on the left side of the initial discontinuity for the 3rd model (Fig. 13 only) with respect to the initial temperature of previous models. Thus, we show only the abundance distributions relative to models in which only the left-side initial temperatures are different: the 2nd model, whose initial temperature ratio left-right is equal to $11$ (Fig. 12) as in the 1st model previously discussed, together with the 3rd model results, whose initial temperature ratio left-right is $22$ (Fig. 13).

\begin{figure}
\resizebox{\hsize}{!}{\includegraphics[clip=true]{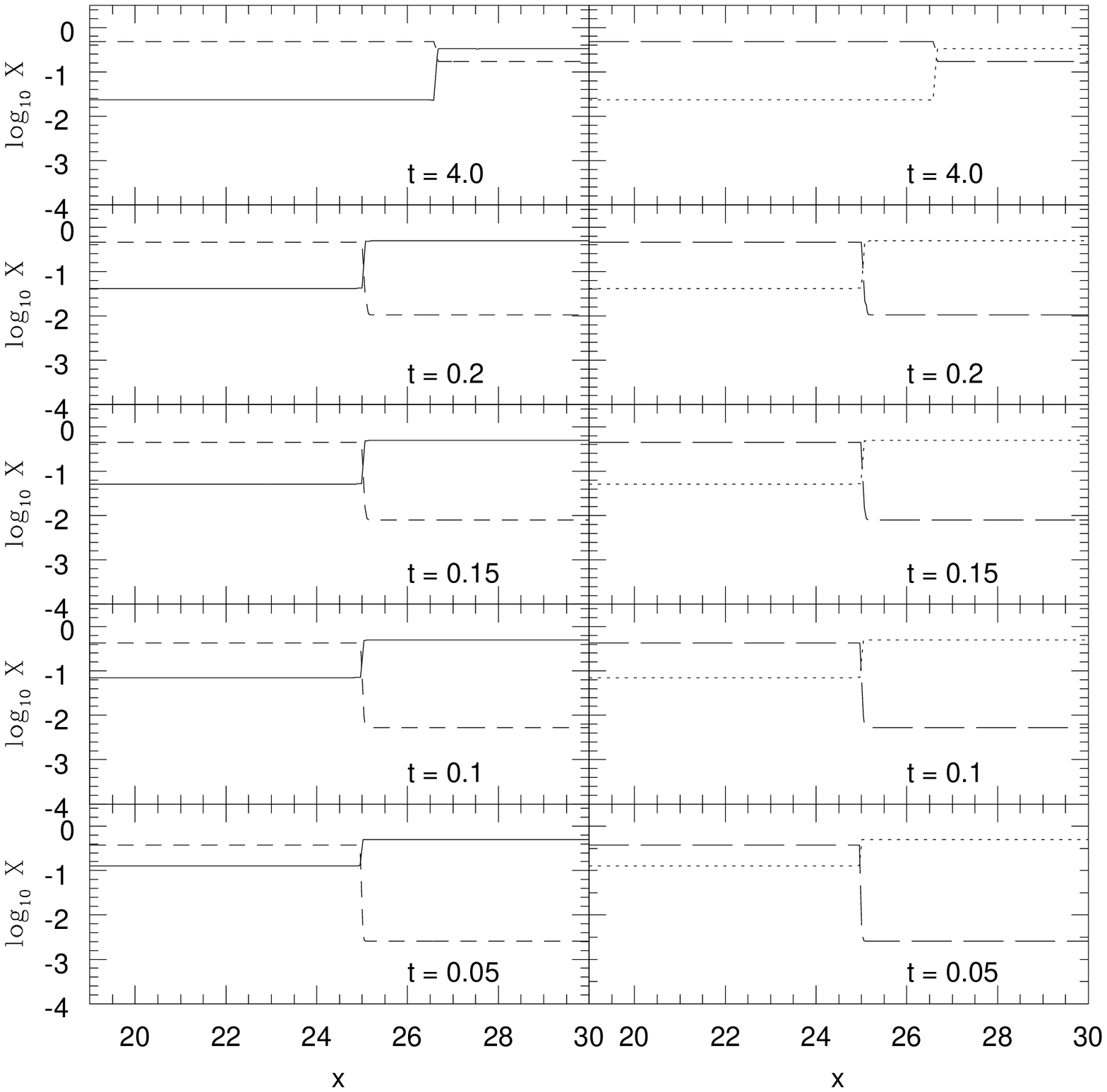}}
\caption{4th model abundance distributions of $C$, $CO$, $O_{2}$ and $O$, triplicating the initial temperature of the left side discontinuity flow with respect to that relative to Fig. 11.}
\end{figure}

\begin{figure}
\resizebox{\hsize}{!}{\includegraphics[clip=true]{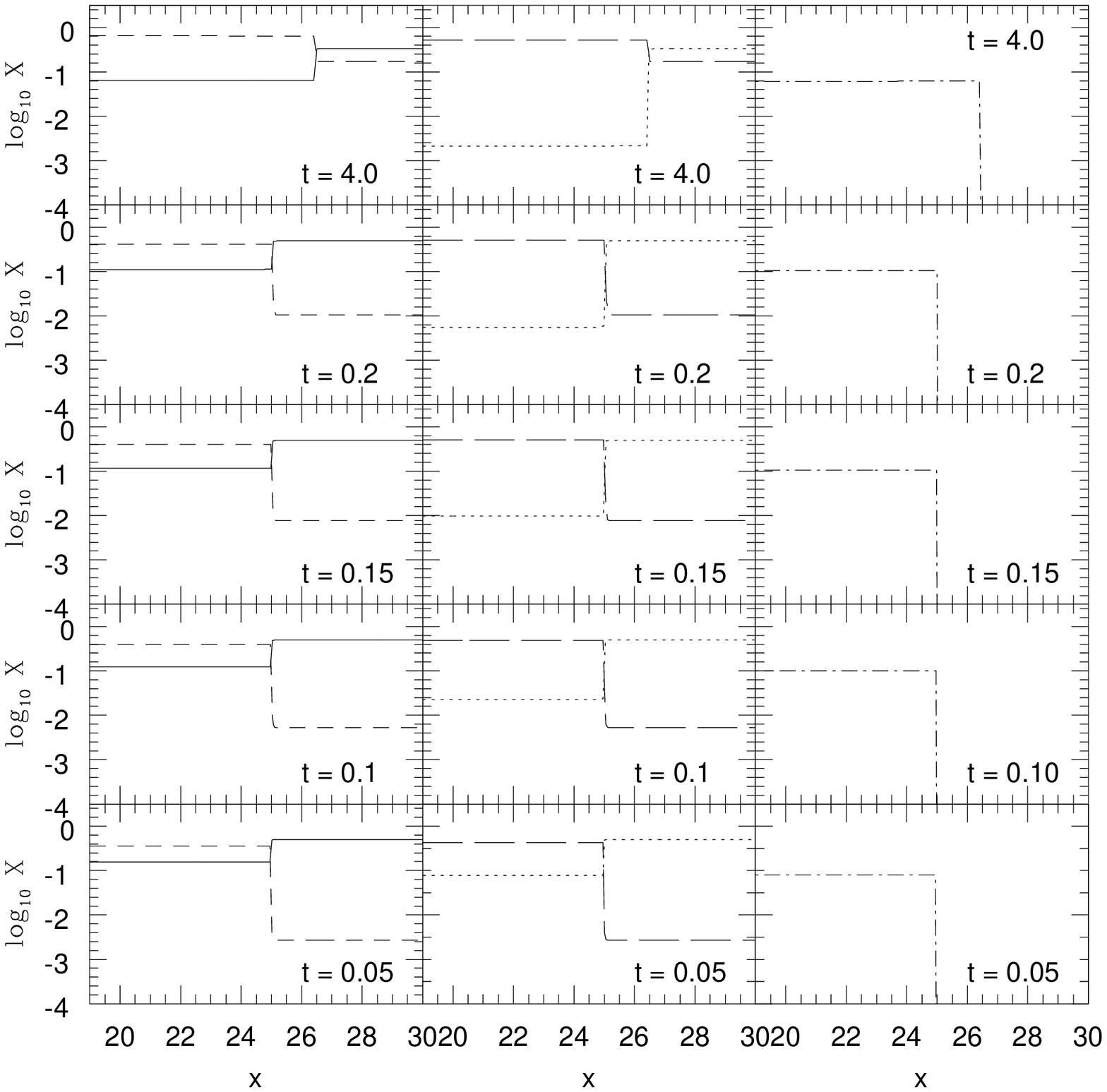}}
\caption{5th model abundance distributions of $C$, $CO$, $O_{2}$, $O$ and $CO_{2}$, triplicating the initial temperature of the left side discontinuity flow with respect to that relative to Fig. 12.}
\end{figure}

  The two reactive and nonreactive discontinuity flow profiles still show differences only in the ($2 \rightarrow 1$) scalability in the profile of $\epsilon$ (here not shown), since the abundance of the $CO_{2}$ is within $\approx 1\% - 6\%$ at the saturation level ($t = 4$ in both figures 12 and 13), so that, any contribution of $CO_{2}$ on the mean molecular weight, gas compressibility and thermal energy per unit mass is small in both cases. Besides, any further temperature variations, due to the reactions involving the $CO_{2}$ and the $CO + O \rightarrow O_{2} + C$, are still negligible. However, what is relevant is that even a so tiny molecular formation is also catched, even working with a simple precision code version.

  Differences in the discontinuity flow profiles are, instead, shown by triplicating the initial temperature on the left side of the initial discontinuity (Figs. 14, 15 and 16) for all four reactive $O_{2} + C \leftrightarrows O + CO$ and $CO + O_{2} \leftrightarrows CO_{2} + O$ flows (4th model). As these figures clearly show, the reactive planar profile for $\epsilon$ is intermediate in the hotter left side front of the discontinuity flow between the reactive discontinuity flow profile, obtained considering only the two reactions $O_{2} + C \leftrightarrows O + CO$ reactive flow (here our 5th model), and that relative to the nonreactive flow at the same initial conditions. In these hotter conditions, the combination of a tiny cooling, due to both the reactions involving both the $CO_{2}$ and the $O + CO \rightarrow O_{2} + C$, as well as the $\gamma$ decrease and the contemporary increase of $\overline{\mu}$, give a decrease in the reactive flow heating (Fig. 15), as well as a decrease in the $\overline{\mu}^{-1} (\gamma - 1)$ term in the EoS and consequently a decrease of pressure. From the molecular abundance distributions shown in Figs. 17 and 18, at time $t = 4$, $\overline{\mu} \approx 22$ and $\gamma \approx 1.53$ for the 5th model, while $\overline{\mu} \approx 24.2$ and $\gamma \approx 1.46$ for the 4th model, respectively. That means a reduction of the $\overline{\mu}^{-1} (\gamma - 1)$ term in the EoS from $\approx 2.4 \cdot 10^{-2}$ (5th model) to $\approx 1.5 \cdot 10^{-2}$ (4th model). This explains why, as shown in Fig. 16, the planar profiles of the velocity $v_{x}$ for the full reactive flow, where all 4 reactions are considered, are consequently less prominent, even compared with those relative to the nonreactive flow model.

  Comparing the planar discontinuity flow profiles of the thermal energy per unit mass for reactive flows, dealing only with $O_{2} + C \leftrightarrows O + CO$ reactions (Figs. 9 and 15), it is noteworthy that the $\Delta \epsilon$ discrepancy between the left side reactive front $\epsilon$, to the respective nonreactive counterpart is the same in both figures, whatever is the time $t$ from the saturation-equilibrium onwards. This is another witness of the quality of calculations, because although the fraction of thermal energy power per unit mass, coming from combustion reactions, as well as the molecular reaction rates, both strongly depend on the local temperature and on the reactant densities, once the equilibrium is established at saturation, any extension in the time integration gives a null reactive contribution to $\epsilon$. Therefore, all the integrated energy per unit mass, released by combustion reactions, is correctly only a function of the initial reactant densities. This direct comparison is made possible since, for the two reactions $O_{2} + C \leftrightarrows O + CO$, both $\overline{\mu}$ and $\gamma$ in the EoS conserve their initial values throughout.

\begin{figure}
\resizebox{\hsize}{!}{\includegraphics[clip=true]{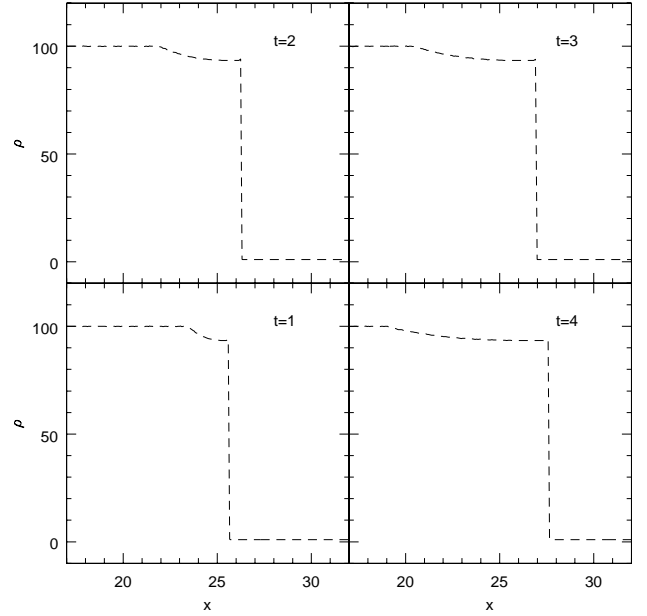}}
\caption{6th model planar discontinuity flow profiles of total mass density $\rho$ at times $t = 1, 2, 3, 4$. The two profiles: reactive and nonreactive overlay with each other. The nonreactive profile is shown in a dashed line.}
\end{figure}

\begin{figure}
\resizebox{\hsize}{!}{\includegraphics[clip=true]{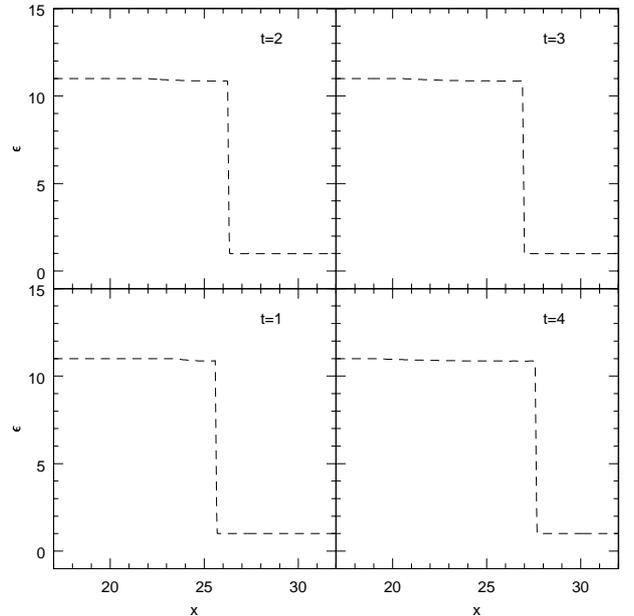}}
\caption{6th model planar discontinuity flow profile of energy per unit mass $\epsilon$ at times $t = 1, 2, 3, 4$. The two profiles: reactive and nonreactive overlay with each other. The nonreactive profile is shown in a dashed line.}
\end{figure}

\begin{figure}
\resizebox{\hsize}{!}{\includegraphics[clip=true]{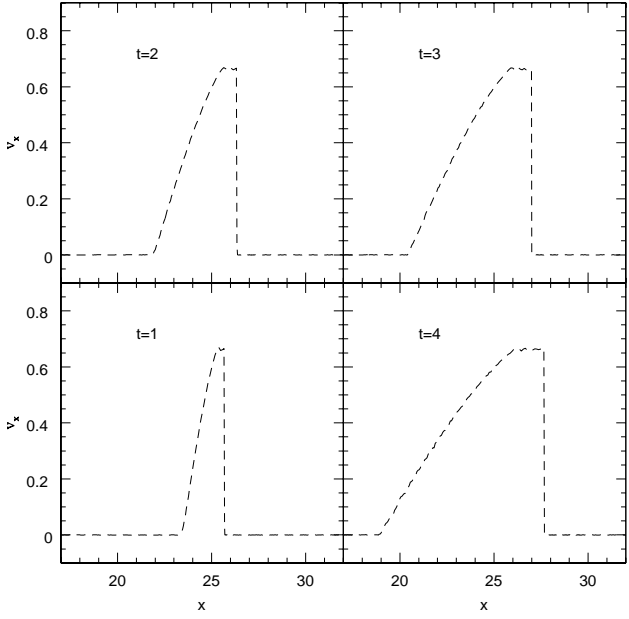}}
\caption{6th model planar discontinuity flow profile of velocity $v_{x}$ at times $t = 1, 2, 3, 4$. The two profiles: reactive and nonreactive overlay with each other. The nonreactive profile is shown in a dashed line.}
\end{figure}

  It is worth noting that, as shown in these figures, dealing in particular with the $CO_{2}$ abundance distributions, any abundance distribution saturation level is obtained in a waving form, towards the final tract of the saturation path. This is evidenced by the small decrease of the $CO_{2}$ abundance distribution of Fig. 18 at time $t = 4$, compared with its values around $t \simeq 0.2$, being its peak value ($\approx 12\% - 15\%$) around $t \approx 1$, corresponding to the maximum energy deviation (decrease), as shown in the four enlargements in Fig. 15.

\begin{figure}
\resizebox{\hsize}{!}{\includegraphics[clip=true]{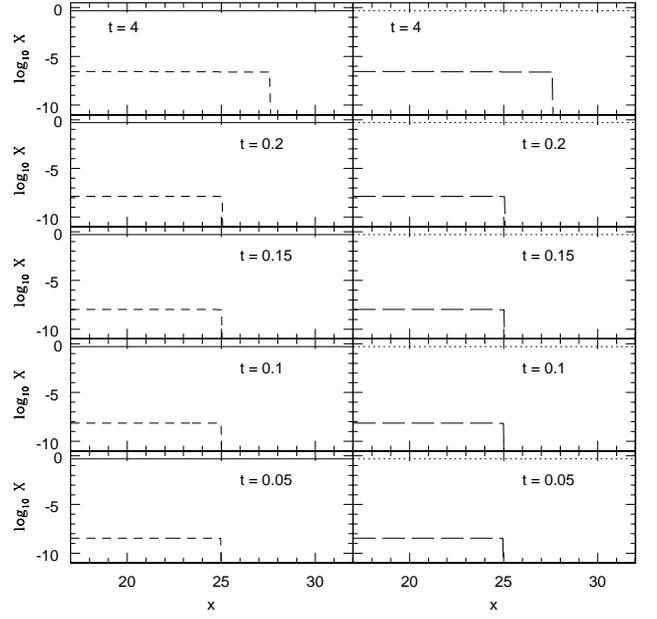}}
\caption{6th model abundance distributions of $^{6}Li$ in solid line, and $^{3}He$ in short dashed line on the left panels of the plot at times $t = 0.05, 0.1, 0.15, 0.2, 4$. Abundance distributions of $^{1}H$ in dots, and $^{4}He$ in long dashed line on the right panels of the plot at the same instances.}
\end{figure}

\subsection{Nuclear chemistry on 2D planar discontinuity flows: Reactive and nonreactive flow computational tests}

  In the 6th numerical experiment performed, dealing with nuclear reactive discontinuity flow flows, we only consider the reactions: $^{6}Li + ^{1}H \leftrightarrows ^{3}He + ^{4}He$ at the moment, where the initial abundances of $^{6}Li$ and protons are both $0.5$ on both sides of the initial discontinuity. The initial temperature on the right side of the initial discontinuity is $10^{7}$ K. The denser left side of the initial discontinuity is $11$ times hotter and the state equation of ideal gases is still adopted in such conditions.

  In nuclear chemistry, since all nuclides are monoparticles, $\gamma = 5/3$ throughout. Making things simpler, even the mean molecular weight $\overline{\mu}$ is unchanged for the reactive flow we are considering.

  Of the two nuclear reactions: $^{6}Li + ^{1}H \rightarrow ^{3}He + ^{4}He$ and $^{3}He + ^{4}He \rightarrow ^{6}Li + ^{1}H$, the first one, exothermic, prevails since the two rate coefficients \citep{c31} are:

\begin{eqnarray}
C_{^{6}Li + p \rightarrow ^{3}He + ^{4}He}(T) & = & \frac{3.7 \cdot 10^{10}}{T_{9}^{2/3}} e^{- \frac{8.413}{T_{9}^{1/3}} - \left( \frac{T_{9}}{5.5} \right)^{2}} \cdot \nonumber \\
& & \cdot \big( 1. + .05 T_{9}^{1/3} - \nonumber \\
& & - .061 T_{9}^{2/3} + .021 T_{9} + \nonumber \\
& & .006 T_{9}^{4/3} + .005 T_{9}^{5/3} \big) + \nonumber \\
& & \frac{1.33 \cdot 10^{10}}{T_{9}^{3/2}} e^{- \frac{17.763}{T_{9}}} + \nonumber \\
& & \frac{1.29 \cdot 10^{9}}{T_{9}} e^{- \frac{21.02}{T_{9}}}, \\
C_{^{3}He + ^{4}He \rightarrow ^{6}Li + p}(T) & = & C_{^{6}Li + p \rightarrow ^{3}He + ^{4}He}(T) \cdot \nonumber \\
& & 1.07 e^{- \frac{46.653}{T_{9}}},
\end{eqnarray}

where $T_{9} = 10^{-9} T$, differ by the term $1.07 e^{- 46.653/T_{9}}$, strongly reducing the rate of $^{3}He + ^{4}He \rightarrow ^{6}Li + ^{1}H$ reaction with respect to the $^{6}Li + ^{1}H \rightarrow ^{3}He + ^{4}He$.

  Figures 19, 20 and 21 show the planar profiles of the mass density $\rho$, of thermal energy per unit mass $\epsilon$ and of velocity $v_{x}$ only for nonreactive flows. The rate coefficients are very small in the temperature range considered, so that any difference between the reactive and the nonreactive planar profiles for $\rho$, $\epsilon$ and $v_{x}$, come out very slowly, over time. Thus, any difference is not yet evidenced in the four panels of these figures, since the adiabatic accumulation of reactive energy is too slow. The abundance distributions of the produced nuclides are shown in Fig. 22, where the planar profiles of chemicals are reported for four selected instants, showing both their initial progression and their final setting. In Fig. 22, it is shown that both $^{3}He$ and $^{4}He$ abundances slowly increase up to $\approx 10^{-7} - 10^{-6}$ at time $t = 4$, that is still very far from the equilibrium configuration because the $^{6}Li + ^{1}H \leftarrow ^{3}He + ^{4}He$ works very slowly, and the reverse reaction is practically ineffective in so far as $T \sim 10^{8} - 10^{9}$ K, it is peeping at $T \sim 10^{9} - 10^{10}$ K, and it is concurrent at $T \sim 10^{10} - 10^{11}$ K.

  In both reactive and unreactive planar discontinuity flow profiles for velocity $v_{x}$, very small evidences of Gibbs phenomenon \citep{c40,c41} are visible on the top plateau of the square wave $v_{x}$ profile. Moreover, tiny ripples are also visible on the same flat zone on the top of $v_{x}$, where the nonreactive analytical solution profile is flat, while they are absent on the planar profile of $\epsilon$ because of the more effective smoothing out of the artificial smoothing out numerical terms in the energy equation than in the momentum equation. These ripples on the $v_{x}$ profiles, although less evident, were also found in the previous moleculare chemistry discontinuity flow tests. The handling of such ripples and their smoothing out, are argument better concerning the best choice that should be made to the artificial dissipation terms (here $\alpha_{SPH} = 1$ and $\beta_{SPH} = 2$) introduced in the momentum and in the energy equations \citep{c38} and where and how they should locally be greater or smaller, a matter that is beyond the scope of this paper.

\begin{figure}
\resizebox{\hsize}{!}{\includegraphics[clip=true]{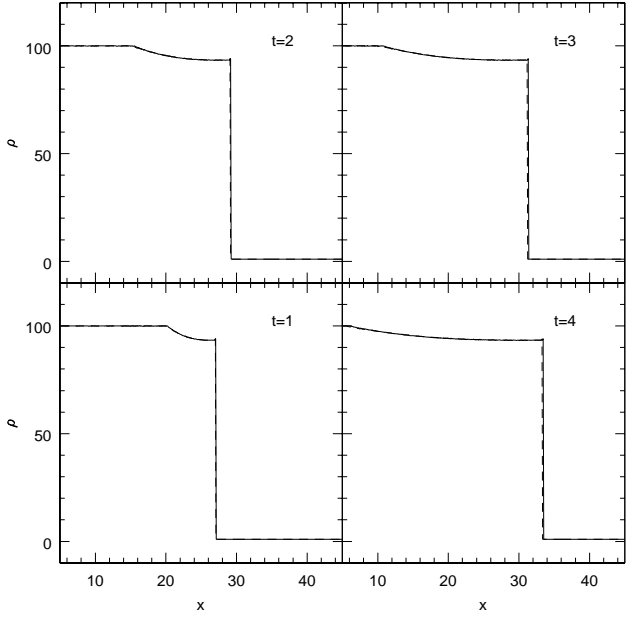}}
\caption{7th model planar discontinuity flow profile of total mass density $\rho$ at times $t = 1, 2, 3, 4$ for reactive flows where $^{6}Li + ^{1}H \leftrightarrows ^{3}He + ^{4}He$ reactions occur. The unreactive discontinuity flow profile is also shown for comparison in a dashed line profile.}
\end{figure}

\begin{figure}
\resizebox{\hsize}{!}{\includegraphics[clip=true]{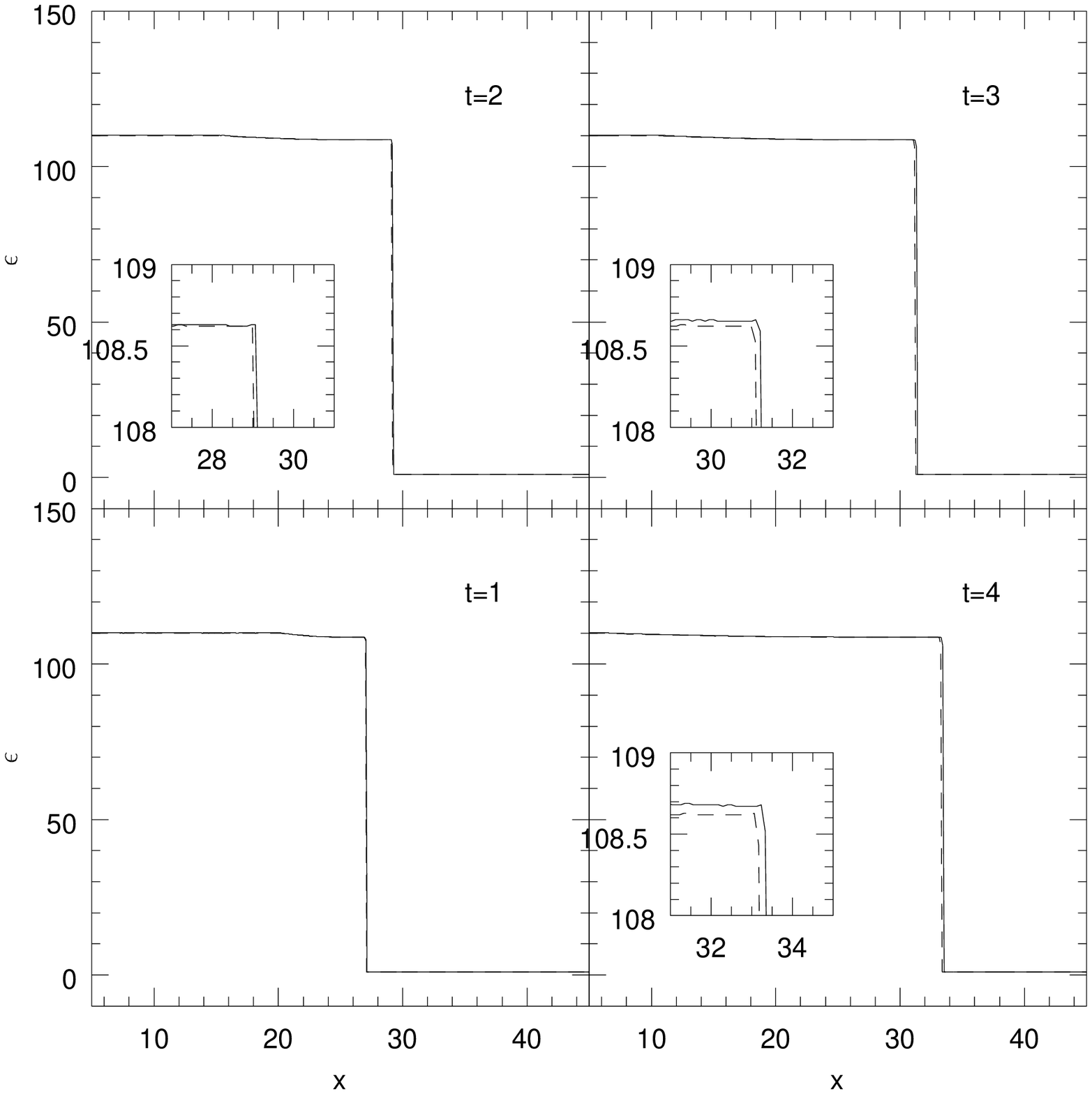}}
\caption{7th model planar discontinuity flow profile of thermal energy per unit mass $\epsilon$ at times $t = 1, 2, 3, 4$ for reactive flows where $^{6}Li + ^{1}H \leftrightarrows ^{3}He + ^{4}He$ reactions occur. The unreactive discontinuity flow profile is also shown for comparison in a dashed line profile. Differences are better shown in some panels in dedicated enlargements.}
\end{figure}

\begin{figure}
\resizebox{\hsize}{!}{\includegraphics[clip=true]{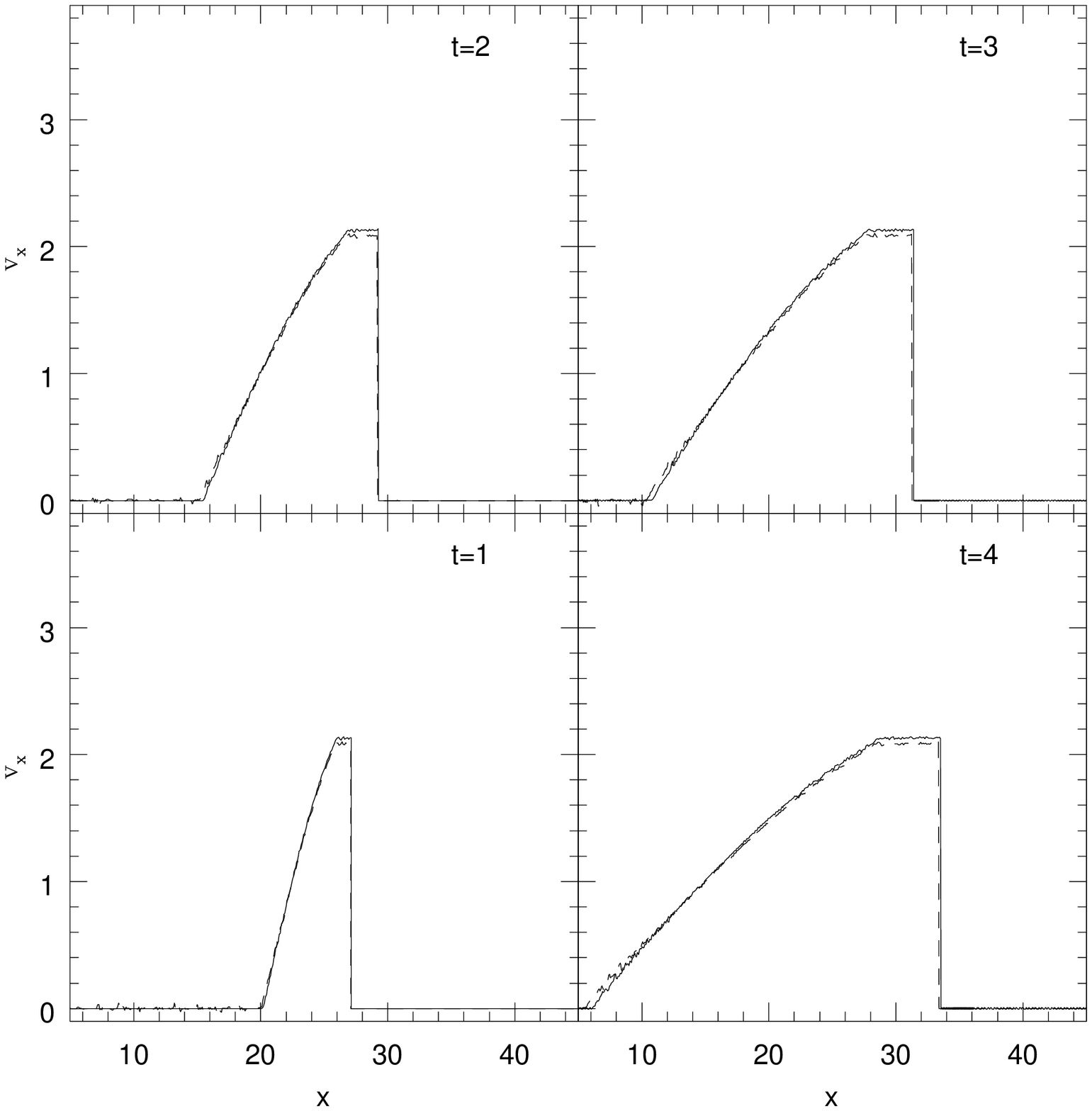}}
\caption{7th model planar discontinuity flow profile of velocity $v_{x}$ at times $t = 1, 2, 3, 4$ for reactive flows where $^{6}Li + ^{1}H \leftrightarrows ^{3}He + ^{4}He$ reactions occur. The unreactive discontinuity flow profile is also shown for comparison in a dashed line profile.}
\end{figure}

\begin{figure}
\resizebox{\hsize}{!}{\includegraphics[clip=true]{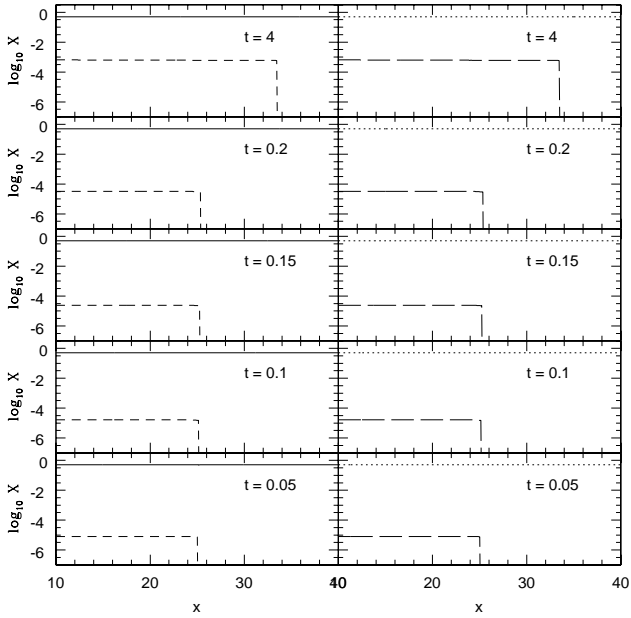}}
\caption{7th model abundance distributions of $^{6}Li$ in solid line, and $^{3}He$ in short dashed line on the left panels of the plot at times $t = 0.05, 0.1, 0.15, 0.2, 4$. Abundance distributions of $^{1}H$ in dots, and $^{4}He$ in long dashed line on the right panels of the plot at the same instances.}
\end{figure}

  In the 7th numerical model, still considering the same initial conditions for $v_{x}$, $\rho$, particle masses, for the same two nuclear reactions previously considered in the 6th model, we assumed a left side front still hotter, where the temperature ratio between the two left-right sides of the initial discontinuity is equal to $110$. In so doing, the reaction rate coefficient, as expressed by (eq. 82) is $\approx 4 \cdot 10^{3}$ times larger than in the 6th previous model. As it is shown from Fig. 23 to Fig. 25, the evolution of the reactive discontinuity flow front shows some slight differences with respect to the pertaining unreactive modelling because the accumulation energy released by reactions starts to be appreciable.

  In such 7th last numerical test, small evidences for Gibbs phenomenon \citep{c40,c41} are still visible on the top of the square wave throughot the 4 panels in the plots of the velocity $v_{x}$, for both nonreactive and for reactive flows. These characteristics also affect, although much less, especially the right corner to the respective planar profile of $\rho$ (Figs. 19 and 23). The relative heights of these right corner humps are roughly time independent throughout the entire evolution of the discontinuity flow front, in so far as the height of the square wave does not change.

  Figure 26 shows the abundances of all four nuclides $^{6}Li$, $^{1}H$, $^{3}He$ and $^{4}He$ for four selected instants, showing both their initial progression and their final setting. Despite the chemical abundances are still far from any equilibrium-saturation configuration, the increase of both helium abundances up to $10^{-3}$ are peeping in terms of nuclear energy released from $t = 4$ onwards. If we consider that $\approx 10^{8}$ binuclear reactions involving $\approx 4$ MeV each occur in a fluid having $\rho = 10^{-12}$ g cm$^{-3}$, being the normalization value for the thermal energy per unit mass $\epsilon_{\circ} \approx 1.2 \cdot 10^{15}$ erg g$^{-1}$, we get an energy per unit mass deviation reactive-to-unreactive $\Delta \epsilon|_{reac} \approx 0.25$, a value that is not far from the separation from the two energy profiles, reactive to unreactive, shown in the enlargement in the fourth panel of figure 24.

  Thus, these results made we aware that the goal of an equilibrium configuration of nuclear abundances would imply a much longer $x$ line, working with mass densities of the order of magnitude used in the last two models here shown (6th and 7th). Therefore, the goal of the equilibrium would imply a much greater number of particles, much larger arrays and a much longer computational time, even considering that, at these high temperatue values, the computational time step decreases of one order of magnitude with respect to that of the previous 6th model. This compels us to reduce the comparison among reactive to nonreactive flow discontinuities as far as it is possible up to a certain time limit, because the order of magnitude of energy ranges are very different.

\begin{figure}
\resizebox{\hsize}{!}{\includegraphics[clip=true]{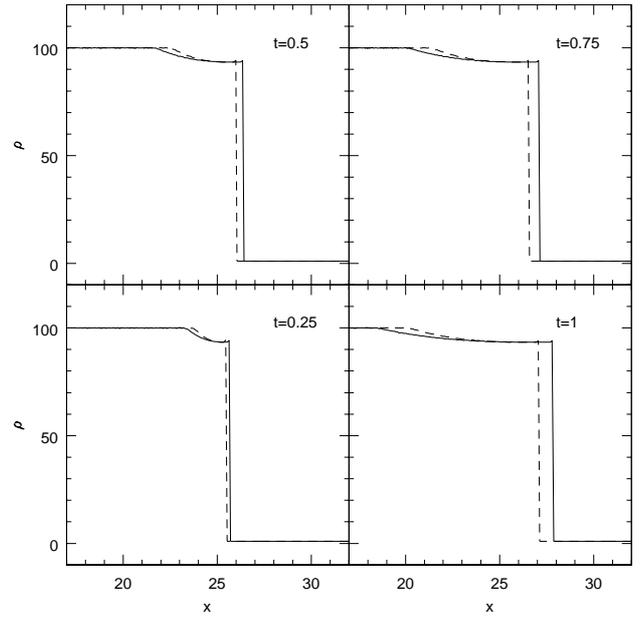}}
\caption{8th model planar discontinuity flow profile of total mass density $\rho$ at times $t = 0.25, 0.5, 0.75, 1$ for reactive flows where $^{6}Li + ^{1}H \leftrightarrows ^{3}He + ^{4}He$ reactions occur. The unreactive discontinuity flow profile is also shown for comparison in a dashed line profile.}
\end{figure}

\begin{figure}
\resizebox{\hsize}{!}{\includegraphics[clip=true]{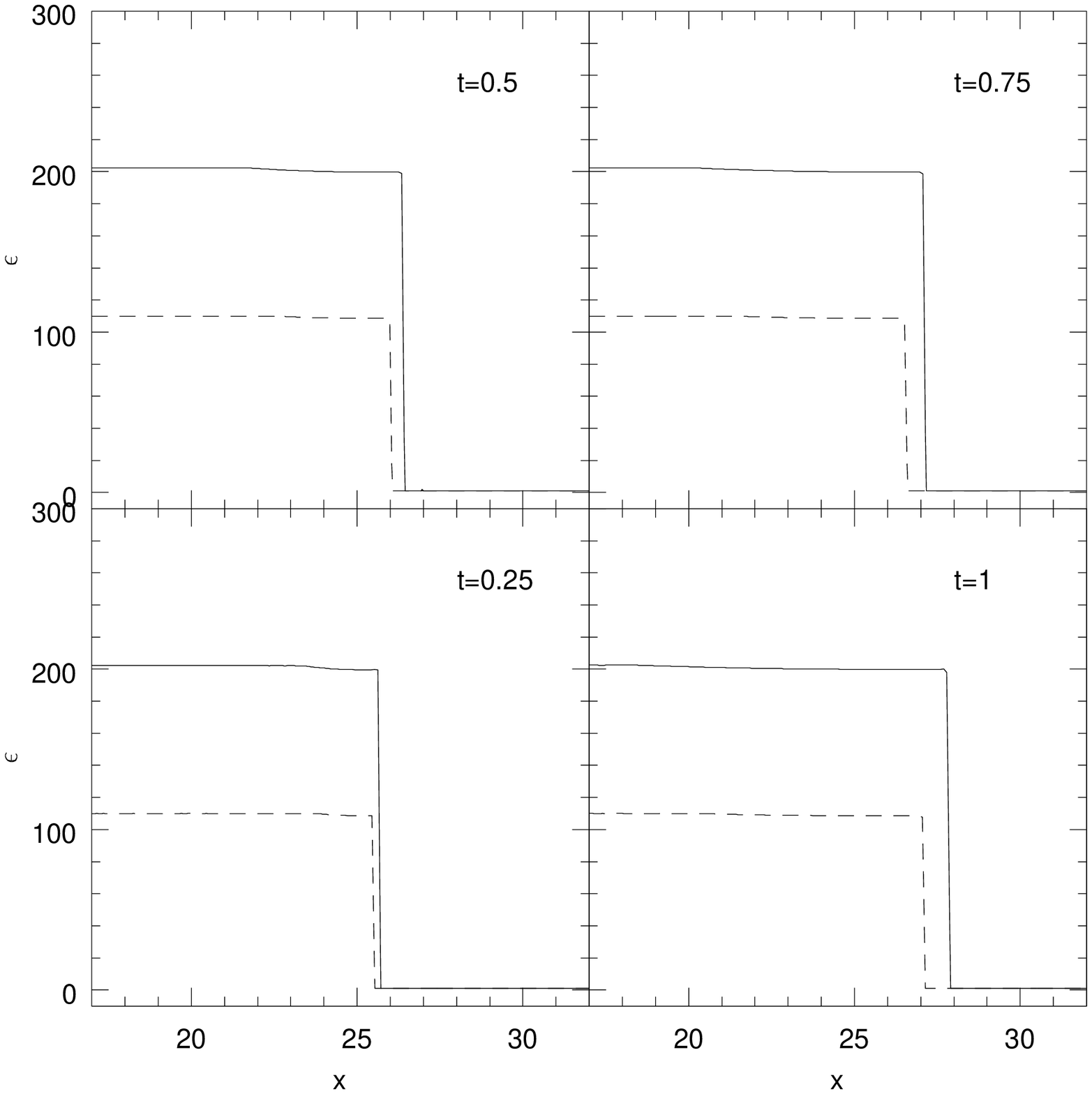}}
\caption{8th model planar discontinuity flow profile of thermal energy per unit mass $\epsilon$ at times $t = 0.25, 0.5, 0.75, 1$ for reactive flows where $^{6}Li + ^{1}H \leftrightarrows ^{3}He + ^{4}He$ reactions occur. The unreactive discontinuity flow profile is also shown for comparison in a dashed line profile. Differences are better shown in some panels in dedicated enlargements.}
\end{figure}

\begin{figure}
\resizebox{\hsize}{!}{\includegraphics[clip=true]{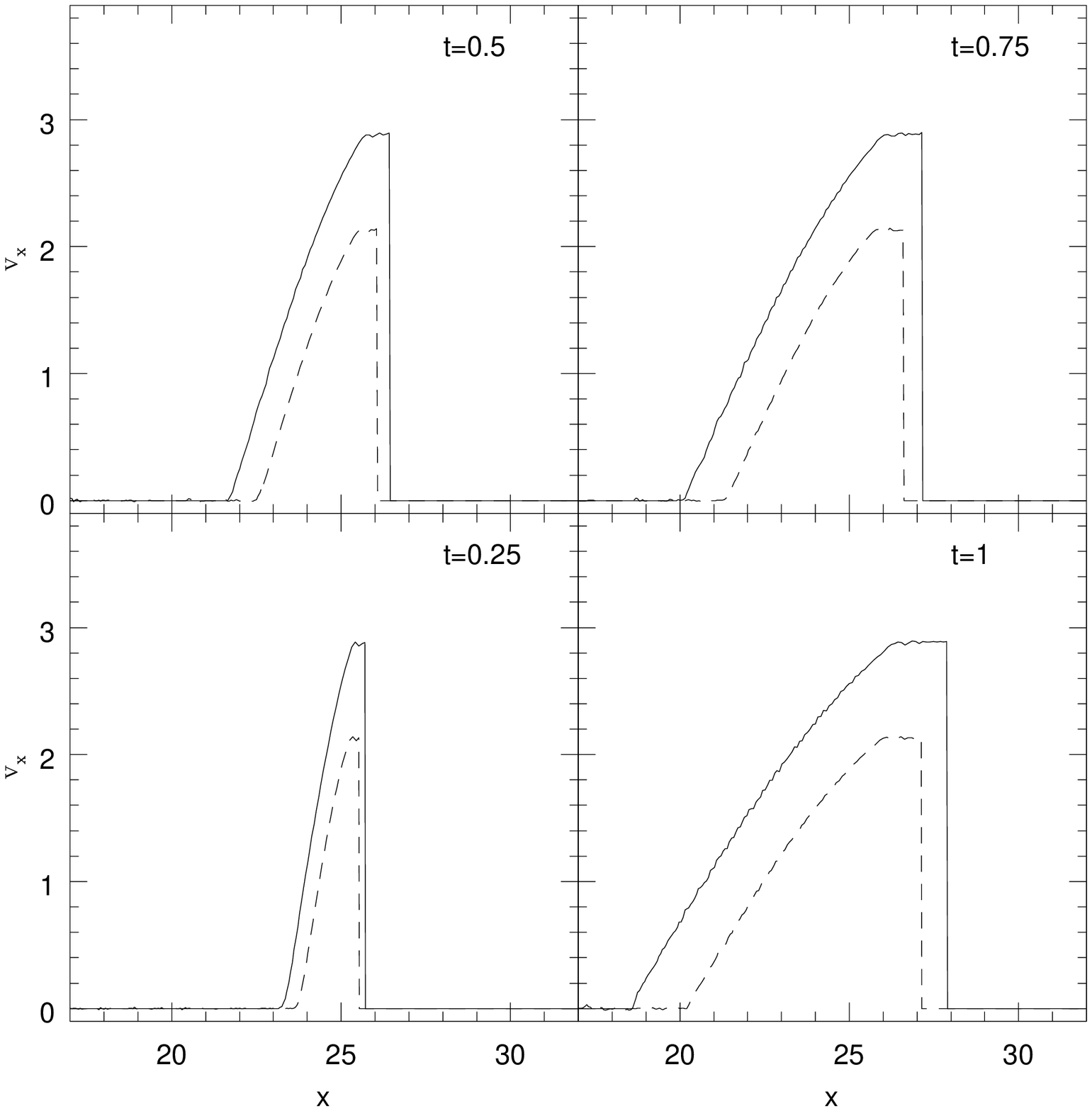}}
\caption{8th model planar discontinuity flow profile of velocity $v_{x}$ at times $t = 0.25, 0.5, 0.75, 1$ for reactive flows where $^{6}Li + ^{1}H \leftrightarrows ^{3}He + ^{4}He$ reactions occur. The unreactive discontinuity flow profile is also shown for comparison in a dashed line profile.}
\end{figure}

\begin{figure}
\resizebox{\hsize}{!}{\includegraphics[clip=true]{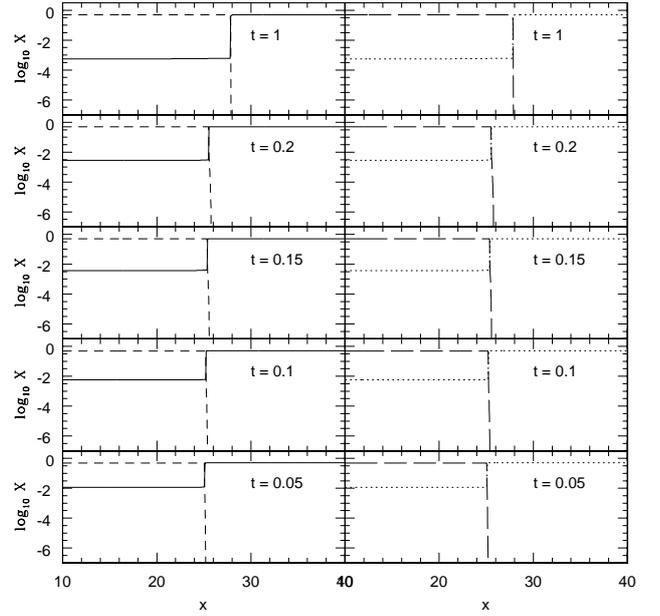}}
\caption{8th model abundance distributions of $^{6}Li$ in solid line, and $^{3}He$ in short dashed line on the left panels of the plot at times $t = 0.05, 0.1, 0.15, 0.2, 1$. Abundance distributions of $^{1}H$ in dots, and $^{4}He$ in long dashed line on the right panels of the plot at the same instances.}
\end{figure}

  Since, according to eq. (66) the reaction rate is strongly dependent on the mass density of reactants, we also performed a 8th model of neclear reactive flow modelling, based on the same initial conditions as the 7th previous model, but using a normalization density $\rho_{\circ} = 10^{-8}$ g cm$^{-3}$ instead of $\rho_{\circ} = 10^{-14}$ g cm$^{-3}$. This choice also reflects on the thermal energy power per unit mass because of eq. (9).

  Figs. 27 to 29 show results for $\rho$, $\epsilon$ and $v_{x}$ at four selected instants each, while Fig. 30 shows the abundances for all four nuclides involved in the two reactions we considered. As it is evident, much bigger differences in the reactive to unreactive flow comparison come out, while secondary details on the flow profiles have yet been discussed in the previous models, so that we do not repeat them again. But, how it is clearly shown in these pictures and what is very important is that most of the burning of the initial reactants occurred in the first instants, leading to an abrupt fluid temperature rise on the left side of the flow initial discontinuity, getting its equilibrium-saturation level, where the leading reaction is amolst exhausted, since the decrease of both $^{6}Li$ and $^{1}H$ sets up around $10^{-3}$, a value still far from $\approx 10^{-10}$, corresponding to the ratio of the reaction rate coefficients at $T \approx 10^{9}$ K, but low enough to consider the exaustion of reactants. To conclude, repeating the previous estimation, made before for the 7th model, if we consider for both helium isotopes that $\approx 2 \cdot 10^{17}$ binuclear reactions involving $\approx 4$ MeV each occur in a fluid having $\rho = 10^{-6}$ g cm$^{-3}$, being the normalization value for the thermal energy per unit mass $\epsilon_{\circ} \approx 1.2 \cdot 10^{15}$ erg g$^{-1}$, we get a $\Delta \epsilon|_{reac} \approx 110$, very close to that shown throughout the four panels of Fig. 28.

\subsection{Some remarks on the basic hypotheses}

  As a first remark, in spite of the fact that the chemistry tool is free of any adopted numerical scheme, we explain why we prefer working with a Gaussian-related Kernel, using a steady spatial resolution length $h$, instead of using an adaptive SPH.

  First of all, there are all motivations widely discussed in \citep{c10}, but there is also a physical reason, dealing with the numerical interpolation accuracy.

  Working with reactive flows, any fluid spatial component should be strictly considered much more tightly as a real fluid portion than as a flow component. In so doing, we stick closely follow what it is explicitly written in \citep{c9}: "if you want to find a physical interpretation of an SPH equation, it is always to assume the kernel is a Gaussian. This is the first golden rule of SPH''.

  In SPH, the Kernel spatial derivatives are necessary for the computation of the unreactive component of Lagrangian time derivatives of $\rho$, $\epsilon$ and $\mathbf{v}$ through a spatial interpolation-integration process, followed by a temporal integration procedure. In the adaptive SPH, it is customary the adoption of some averaging. Whenever $h_{i} \neq h_{j}$: either

\begin{equation}
W_{ij} = W(|\mathbf{r}_{ij}|, h_{ij}),
\end{equation}

where
\begin{equation}
h_{ij} = 0.5 \cdot (h_{i} + h_{j})
\end{equation}

\citep{c42}, or

\begin{equation}
W_{ij} = 0.5 \cdot (W(|\mathbf{r}_{ij}|, h_{i}) + W(|\mathbf{r}_{ij}|, h_{j}))
\end{equation}

\citep{c43} are adopted to compute the spatial derivatives of the unreactive component of the system of Euler equations (eqs. 11-15) through their conversion in the SPH formalism. As eqs. (12-14) clearly show, the spatial resolution length $h$ has a role in the SPH conversion of the unreactive component of the system of Euler equations through the adopted interpolation Kernel, as well as it also locally affects the necessary artificial dissipation playing a role in the particle collisions \citep{c9,c12}. Instead, the two reactive summations added to the right side of the SPH conversion equations (eqs. 12, 14) are free of any dependence on $h$. This occurs because the characteristic speed of reactions $v_{reac,i}$ is physically intrinsecally independent on any characteristic spatial length, being related to the frequency of molecular collisions. However, $v_{reac,i}$ is necessary to the computation of $\Delta t_{SPH}^{CFL}$ (eqs. 76, 78), where only $h_{i}$ is used.

  Being anyway far from the continuum, Kernel spatial derivatives for the same particle configuration on the same $i$th particle, computed using any averaging, differ from each other as well as they differ from that computed working with a steady $h_{i}$. This occurs as much as farther from the continuum limit is the distribution of particle companions. This is unavoidably due to the fact that interpolation errors in SPH not only depend on the number of neighbours $N$ (being $\propto N^{-1/2}$), but errors also intrinsecally strongly depend both on $h$ as well as on the neighbour spatial distribution \citep{c44,c45,c46}. In principle, for a Kernel profile free of any Gaussian relationship, it is not sure whether the greater is $N$, the smaller are the interpolation-integration errors because $N \rightarrow \infty$ as $h \rightarrow \infty$, while errors should go to zero even as $N \rightarrow \infty$ while $h \rightarrow 0$ in the continuum limit. Thus, the local reactive thermodynamics in adaptive SPH, could lead to as many differences in the abundances of chemicals as many adaptive criteria are considered, and consequently to as many additional differences in the local thermodynamics over time through the chemical heating, as well as through $\gamma$ and $\overline{\mu}$ if reactions change their initial values.

  This problem occurs in particular whenever the reaction rate coefficient (eq. 63) is a steep function of temperature, without neglecting the role of the number density of reacting chemicals on reaction rates (eq. 65), so that $v_{reac,i}$ deeply shorten the time steps. Any difference, using different reactive SPH approaches, should be catched well before any equilibrium configuration. In fact, differences in the abundances of chemicals exist throughout the intermediate configurations, also reflecting on the hydrodynamics, because of different thermodynamics dictated by the chemical speeds. Perhaps, but this needs a further dedicated feasibility study, an expression for $\Delta t_{SPH}^{CFL}$ (eq. 78) could better mitigate this difficulty by using $h_{ij}$ instead of $h_{i}$ through another algorithm, explicitly introducing some dependence on the spatial resolution length for pairs of contiguous fluid components, even on the reactive components of eqs. (1,3). Instead, whenever the reaction rate coefficients (eq. 63) are almost independent on temperature especially in rarefied diffuse astrophysical environments and, above all, $\Delta t_{SPH}^{CFL} = \Delta t_{SPH}^{reac}$ because of the low speed of reactions with respect to the other characteristic speeds, the SPH formulations using adaptive averagings do not strongly suffer of this chemical instability of transient phases.

  Summing up this first remark we could synthesize that while for unreactive ideal fluid dynamics the computed momentum and energy per unit mass are independent on the fluid mass and mass density, whilst mass and mass density are scaled, for the reactive counterpart that it is not so, as it is shown in the previous sections. Thus, reactive flow calculations could lead to further inaccuracies both in the local chemical abundances and in the local thermodynamics as much as more relevant are the reaction rates. By using SPH, since the accuracy of calculations not only is affected by the number of particle neighbours - in order to reach a satisfactory continuum limit approximation ($N \rightarrow \infty$) - but it also depends on their geometrical distribution \citep{c44,c45,c46}, we prefer adopting the SPH variant \citep{c10} where the function $ W(\mathbf{r}, \mathbf{r}', h) d \mathbf{r}'$ is dimensionally invariant and whose Kernel $W(\mathbf{r}, \mathbf{r}', h)$, coming from the 1D Gaussian profile, is built up using the Error Function. In this fashion, it is as if we were folding the 2D and the 3D interpolation-integration on a single dimension. In so doing, both the particle pairing instability \citep{c10}, affecting any Kernel interpolation profile containing at least an inflection point is solved in 2D and in 3D (not in 1D), as well as any inaccuracy coming out from the geometrical discretization of neighbours which (eqs. 10-11) comes from $r d\theta)$ in 2D and $r^{2} \sin \theta d\theta d\phi$ in 3D in the continuum limit, an inaccuracy which naturally occurs in SPH, even placing the particles in an equal space ordering, in order to set a uniform distribution, coming from the anisotropy of neighbour distribution. Therefore, what stays unsolved is only the inaccuracy in the interpolation-integration coming from the radial inhomogeneity of neighbours.

  As a second remark, throughout the models here shown we assumed some numerical values for the normalization value for the density $\rho_{\circ}$, for the thermal energy per unit mass $\epsilon_{\circ}$ (directly computed from the initial normalization temperature $T_{\circ}$) and $x_{\circ}$ for lengths. Of course, these are arbitrary assumptions dictated by the fact that, being the reaction rate $\rho^{-1} \dot{\rho}_{k}|_{r}$ (eq. 66), a linearly increasing function of $\rho_{\circ}$, the explicit time steps computed through the Courant-Friedrichs-Lewy condition \citep{c35,c36} inavoidably decrease in nonlinear fashion towards unpractical tiny values for larger and larger $\rho_{\circ}$ values. If we need to rescale the computed time steps towards more practical values, it is necessary to act on the other normalization values, in so far it is possible, as it is highlighted in \S 4.3.

  In fact, the full dimensional time step $Dt = t_{\circ} \delta t = \hat{t}_{\circ} \Delta t$, where $\delta t$ is the too small time step, $t_{\circ} = x_{\circ} c^{-1}_{s \circ}$ is the normalization time, $\Delta t$ is the larger time step we need and $\hat{t}_{\circ}$ is the rescaled normalization time. If $\Delta t \gg \delta t$, it is necessary that $\hat{t}_{\circ} \ll t_{\circ}$, that means that this condition is fulfilled by increasing the normalization value (here the sound speed $c_{s \circ}$), and/or by decreasing $x_{\circ}$, whenever it is possible and appropriate to the physical problem we need to study. As an example, reducing $x_{\circ}$ from $10^{5}$ Km to $10^{2}$ cm, we could easily get comparable nondimensional results for a gas dynamics whose $\rho_{\circ}$ is up to $10^{8}$ times larger.

  As a third remark, we remind that those parameters used in the algebraic formulations of rate coefficients are normally updated in new publications over time. This occurs both in the form of fitting parameters in the rate coefficients of algebraic formulations and in the form of fitting tables. Any work in these fields could be soon outdated, especially when compared to that of some decades ago. However, fortunately, since the 80' years, the updating of reaction rates fitting coefficients, or tables, are often second order corrections and new data usually regard only some specific conditions.

  Our fourth and last remark deals with eqs. 66 and 74, dealing with the mass density reaction rates, which we use instead of eq. 72 dealing only with the abundance reaction rates. Both are algebraically and logically correct in so far as we consider molecular chemical reactions, whenever the local mass is strictly conserved, and in so far as we deal with nuclear reactions up to a second order approximation, if mass variations are negligible. However, when the nuclear mass defect plays a role in the mass-energy conservation, the assumption of eq. 66, using the effective mass of nuclides in u.m.a. instead of integer numbers related to the chemical molecular weight of nuclides, gives the exact mass density reaction rate. In this case, if working in SPH, the mass of the $i$th SPH particle should be recalculated each time step as $m_{i} = \rho_{i} n^{-1}_{i}$, being the numerical density $n_{i}$ of the same particle directly obtained by eq. 11 for SPH-like codes, taking care that the smoothing interpolated $n_{i} = \rho_{i} m^{-1}_{i}$ since the starting initial conditions \citep{c39}.

\section{Discussion and conclusions}

  In this paper, we adopted a 2D Lagrangian fluid dynamics, based on the SPH framework \citep{c9}, in its GASPHER variant \citep{c10}, based on a the interpolation integral of the Error function, built up on the basis of the 1D Gaussian profile and modified according whether the simulation is 1D, 2D or 3D. In particular, in 2D and in 3D, the problem of particle pairing instability does not affect 2D and 3D numerical simulations. Any discontinuity at the outer boundary (if truncating) is non-existing, as well as the technique is still a high resolution numerical scheme, as it is well shown by results, where any typical waving behaviour, affecting $O^{2}$ schemes \citep{c20} does not affect $\rho$, $\epsilon$ and $v_{x}$ throughout the models here shown. The $100$ to $1$ density contrast between the left-to-right discontinuity fronts does not show any stability and convergence shortcomings in the interpolation properties. Although the window of spatial interpolation could be extended farther than $2 h$ to get the assigned number of neighbours, any further extension of interpolation is mathematically negligible because the Kernel values are of the order of $10^{-4}$ within $2 h$ and $3 h$ in 1D and even much less in 2D and in 3D.

  Comparisons of analytical solutions with numerical GASPHER results are in very good agreement and shown in \citep{c10}, so that further comparisons here are not necessary. In \citep{c10} all details concerning the accuracy, the momentum and energy conservation, as well as the convergence and numerical stability, have been widely addressed in a wider scenario of cases regarding not only the Riemann problem, but also the mass and momentum transport, and the free edge expansion of fluids.

  In this paper we have shown how a standing alone algorithm, dealing with both molecular and nuclear chemistry, can be included in a Lagrangian fluid dynamic numerical code belonging to the family of SPH \citep{c10}. In principle, this can also be made with other fluid dynamics numerical codes, since local reactive time derivatives, coming from the chemistry module, do not intrinsecally modify the time derivatives of the unreactive fluid dynamic equations or, equivalently, they do not affect the spatial derivatives of the Euler system of equations. Of course, the Courant-Friedrichs-Lewy condition \citep{c35,c36} should also take into account an additional restriction, driven by the reaction rates.

  The module, dealing with reactions, is not conceptually far from that relative to the Eulerian shock tracking FLASH code \citep{c47}, originally developed for thermonuclear reactive processes in novae and in supernovae. However, our code works for a shock capturing Lagrangian fluid dynamics, as well as it also concerns with molecular reactions. Our module, within the Evaluator macros of the numerical scheme of integration, is not conceptually different from that relative to the SPH code written for thermonuclear astrophysics \citep{c48}, with the difference that our code takes into account the full continuity equation (eqs. 1 and 12), instead of calculating the total particle density $\rho$ directly from eq. (11), as it is often made in the majority of SPH codes. Besides, in \citep{c48}, authors solve the equations for the time variation of the chemical abundances $\dot{X}_{k}$, that is the other route of facing this problem (\S 4.2).

  Working with 4-dimensions arrays - for each ith SPH particle we handle three indexes dealing with each chemical species (in turn) and with the two reactants, plus the index of reaction - we also look at the concrete possibility to parallelize at least this new module, according to the hardware and to the software any researcher could set up, apart the simplest OPENMP fortran resource we used. To give an idea on how longer is the time consumption, working with reactive flow chemistry, for two reactions only, involving four chemicals as we did, $80$ iterations are necessary for each $i$th particle in $\{ k, k_{A}, k_{B}, r \}$ nested cycles, where $k = 1, 2, 3, 4$, $k_{A} = 1, 2, 3, 4$, $k_{B} = k_{A}, k_{A} + 1, ..., 4$ and $r = 1, 2$. This does not necessarily imply that the overall computational time spent to do the same number of main integration cycles (see Fig. 2) is $1 + 80/15$ times longer with respect to the unreactive calculation, if $15$ is the assigned number of 2D particle neighbours. However a time consumption $\sim 3 - 4$ longer is not unrealistic, working with a strictly serial programming architecture. By using only the OPENMP resource, we reduce this time for reactive flow calculations up to $\sim 1.3 - 1.5$ times longer than that necessary for nonreactive flow computations. This as for the time consumption for each integration cycle, without considering the tipe step reduction occurring for reactive chemistry flow calculations, which makes longer the computational costs to get the same final integration time.

  In this paper we also performed comparisons of $\rho$, $\epsilon$ and $v_{x}$ for nonreractive planar discontinuity flow profiles, to the respective $\rho$, $\epsilon$ and $v_{x}$ for reactive ones, assuming the same initial thermodynamic and kinetic conditions for a better understanding of both the reactive module efficacy and of the role of thermochemical reactions on the discontinuity flow profiles, and dynamics. This is accomplished through the study of the parameters contained in the EoS: $\epsilon$, $\overline{\mu}$ and $\gamma$. To do this, we have assumed as EoS the equation of state of ideal gases: $p = \overline{\mu}^{-1} (\gamma - 1) \rho \epsilon$, that is the simplest and the more effective expression including these parameters for neutral gases. Even though some specific EoS are adopted for explosive gas detonations, like the Jones-Wilkins-Lee EoS \citep{c49,c50,c51}, even for the SPH method \citep{c50,c51,c52}, the perfect gas EoS can still be adopted \citep{c53,c54,c55}, as it is used for the barionic component of plasmas in the FLASH code \citep{c47}, whose $\gamma = 5/3$. However, for further numerical experiments in much more stressing conditions, other EoS could also be considered as the Redlich-Kwong EoS:

\begin{eqnarray}
p & = & \overline{\mu}^{-1} (\gamma - 1) \rho \epsilon (1 - \rho \overline{\mu}^{-1} N_{A} V_{molec})^{-1} - \nonumber \\ & & a_{RK} T^{-1/2} \rho^{2} \overline{\mu}^{-2} (1 + \rho \overline{\mu}^{-1} N_{A} V_{molec})^{-1},
\end{eqnarray}

where $V_{molec}$ is the mean volume of gas molecules and $a_{RK} \sim 10^{-1} - 10$ l$^{2}$ bar mol$^{-2}$ depends on the specific fluid. This EoS is still used even for denser flows up to the liquid status.

  Examples of some simple planar discontinuity flows have been discussed in this paper when, at the end of the integration time, $\epsilon^{-1}_{unr} \Delta \epsilon|_{reac} < 1$ or it is $\ll 1$, that is when the unreactive thermal energy content per unit mass $\epsilon_{unr}$ to ignite thermochemical reactions is greater or much greater than the variation of the thermal energy per unit mass $\Delta \epsilon|_{reac}$ due to the thermochemical reactions themselves, up to the considered time of integration. This could imply that the saturation-equilibrium configuration could still be not reached if $\epsilon^{-1}_{unr} \Delta \epsilon|_{reac} \gg 1$ in the saturation-equilibrium configuration. Of course, in so far as the flow evolution is computed within restricted characteristic times in nondiffusive LTE adiabatic conditions, a comparison study can hypothetically always be made whatever is $\epsilon^{-1}_{unr} \Delta \epsilon|_{reac}$ in the saturation-equilibrium configuration, even for ratios $\epsilon^{-1}_{unr} \Delta \epsilon|_{reac} \gg 1$, limiting any comparison to models still far from the statistical exhaustion of the reactants. A ratio $\epsilon^{-1}_{unr} \Delta \epsilon|_{reac} \approx 1$, $> 1$ or $\gg 1$, could be a condition fulfilled according to the choice of the overall integration time, respecting the Courant-Friedrichs-Lewy condition (eq. 78), after a huge multitude of time steps, in so far as the combustion reactions are still very far from the equilibrium, after a very long computational time. Examples of still unfavourable thermal conditions when $\epsilon^{-1}_{unr} \Delta \epsilon|_{reac} < 1$ or $\ll 1$, but leading to $\epsilon^{-1}_{unr} \Delta \epsilon|_{reac} \gg 1$ at the saturation-equilibrium after a huge unpractical number of time steps, occur throughout the models here shown on the right side of the flow front discontinuity for the LTE nondiffusive adiabatic models, or whatever is $x$ throughout the last but one 6th model here discussed. Another example deals with stellar cores where an outward energy transport exists, in which the hydrogen burning gives energies $\sim 1 - 10$ MeV at low rates in an environment whose temperature is $\sim 1 - 10$ KeV.

  Of course, the equilibrium could also be obtained for highly exothermic events in nondiffusive adiabatic conditions, according to eq. (9), leading, sooner or later, to reactive flow profiles for $\epsilon$ and $v_{x}$ incomparable to those unreactive. In that case, the risk of occurrence of flow instabilities, whenever the reaction characteristic times are much shorter than the time steps dictated by the unreactive flow Courant-Friedrichs-Lewy condition, is around the corner. 

  Once the equilibrium is reached, the ratio $\epsilon^{-1} \Delta \epsilon|_{reac}$ could be significantly high. Thus, any comparison between reactive and nonreactive discontinuity flows would be an exceptional task since the thermal conditions of adiabatic thermochemical reactive flows are meanwhile totally changed because of the too much integrated energy per unit mass accumulated coming from either from collisional thermochemical reactions, also including the slow neutron capture, or coming from other kinds of exothermic phenomena (e.g. nuclear fission).

  According to the results here shown, nonreactive models are obviously strictly comparable to those relative to thermochemical reactive models whenever the ratio $\epsilon^{-1} \Delta \epsilon|_{reac} \ll 1$ in so far as the roles of $\gamma$ and $\overline{\mu}$ are marginal. Otherwise, clear differences exist, given also by the $\gamma$ and $\overline{\mu}$ variations.

  At the moment, in this paper, we are only interested in showing that the numerical algorithm works well evaluating, at the same time, how the discontinuity flow front evolution and structure are affected by molecular or nuclear reactions. The only way to do this task is to limit the numerical tests only to a few reactions and a few chemicals in nondiffusive adiabatic LTE conditions. Any extension of both should be intended to give answers to some specific problem. Of course, to do this, it is better working with an EoS including all parameters, a particular that does not in most of them because often the EoS are written as fitting empirical formulations. Thus, any further extension (and complication) of input data, regarding any fluid ionization, the plasma EoS, especially for a larger array of chemicals (included electrons) and further reactions are intended to future efforts.

  As it is remarked in \S 5.3, much more numerical pitfalls are hidden for reactive fluid dynamics than for that unreactive. This implies that accuracy of calculations could be a non negligible aspect. If a SPH code is used, then it must be free from any particle pairing instability. This is the main reason because we adopted a modified Gaussian-based SPH Kernel throughout the 2D simulations \citep{c10}, in spite of the fact that even a so manipulated Kernel, satisfaying all mathematical constraints of SPH \citep{c56}, involves a consistency of order 2 (1 at the not discussed boundaries), although numerical solutions of 2D and 3D free edge boundaries are better than in SPH \citep{c10}. Of course, the numerical solutions could be better improved of a further order according to some corrective formulations \citep{c57,c58,c59,c60,c61,c62,c63}, argument that is beyond the scope of this paper at this stage, or it could also be interesting the adoption of other interpolation Kernels \citep{c64} without any inflection points in theirs analytical expressions. However, this efforts deal with further comparisons exploring the capability of further SPH Kernels in approaching with reactive fluid dynamics. At the moment the 2D Kernel we use is correct without any degradation of its adoption to the 1D modelling because the 1D convergency of calculations to the 1D results occur only in the unattainable continuum limit. Thus, it incorrect to affirm that 2D planar flow discontinuity resemble those in 1D.

  To conclude, we need to spend some further words just for those people looking only at the SPH approach. In spite of the fact that some further works have been written \citep{c65,c66,c67,c68,c69,c70,c71} for nuclear chemistry in SPH astrophysical applications, no reactions are explicitly implemented within the system of fluid dynamics equations with the exceptions of \citep{c48} handling a complex array of carbon burning reactions leading to elements of the Fe-Ni-Cr group elements. So, up to date, no strict comparisons to our paper exist, not only including molecular nuclear chemistry, but understanding the role of $\mu$ and $\gamma$ in the EoS by using simple arrays of reactions involving a few chemicals.

{\it Acknowledgements} We are grateful to Dr A.F. Lanza of the INAF - Catania Astrophys. Obs. for the overall helpful discussion, as well as to Dr. K. Weide and Prof. D. Q. Lamb of the Univ. of Chicago for their help on some aspects about the EoS working in the FLASH code.



\end{document}